\documentclass{article}

\usepackage[table]{xcolor}
\usepackage{multirow}
\usepackage{microtype}
\usepackage{graphicx}
\usepackage{subcaption}
\usepackage{booktabs}    
\usepackage{algorithmic}
\usepackage{amsmath}
\usepackage{amssymb}
\usepackage{mathtools}
\usepackage{amsthm}
\usepackage{hyperref} 
\usepackage{xurl}
\usepackage{caption}

\usepackage[preprint]{icml2026}


\usepackage[capitalize,noabbrev]{cleveref}

\theoremstyle{plain}

\theoremstyle{definition}

\theoremstyle{remark}

\newcommand{\nna}{\makebox[\widthof{0.000}][c]{----}}

\usepackage[textsize=tiny]{todonotes}

\icmltitlerunning{ALIEN: Analytic Latent Watermarking for Controllable Generation}

\begin{document}

\twocolumn[
  \icmltitle{ALIEN: Analytic Latent Watermarking for Controllable Generation}

  \icmlsetsymbol{equal}{*}

  \begin{icmlauthorlist}
    \icmlauthor{Liangqi Lei}{bit}
    \icmlauthor{Keke Gai}{bit}
    \icmlauthor{Jing Yu}{muc}
    \icmlauthor{Liehuang Zhu}{bit}
    \icmlauthor{Qi Wu}{uadel}
  \end{icmlauthorlist}

  \icmlaffiliation{bit}{Beijing Institute of Technology, Beijing, China}
  \icmlaffiliation{muc}{School of Information Engineering, Minzu University of China, Beijing, China}
  \icmlaffiliation{uadel}{School of Computer Science, The University of Adelaide, Adelaide, Australia}

  \icmlcorrespondingauthor{Liangqi Lei}{3120245873@bit.edu.cn}
  \icmlcorrespondingauthor{Keke Gai}{gaikeke@bit.edu.cn}

  \icmlkeywords{Machine Learning, ICML, Watermarking}

  \vskip 0.3in
]

\printAffiliationsAndNotice{} 

\begin{abstract}
Watermarking is a technical alternative to safeguarding intellectual property and reducing misuse. 
Existing methods focus on optimizing watermarked latent variables to balance watermark robustness and fidelity, as Latent diffusion models (LDMs) are considered a powerful tool for generative tasks. 
However, reliance on computationally intensive heuristic optimization for iterative signal refinement results in high training overhead and local optima entrapment.
To address these issues, we propose an \underline{A}na\underline{l}ytical Watermark\underline{i}ng Framework for Controllabl\underline{e} Generatio\underline{n} (ALIEN). 
We develop the first analytical derivation of the time-dependent modulation coefficient that guides the diffusion of watermark residuals to achieve controllable watermark embedding pattern.
Experimental results show that ALIEN-Q outperforms the state-of-the-art by 33.1\% across 5 quality metrics, and ALIEN-R demonstrates 14.0\% improved robustness against generative variant and stability threats compared to the state-of-the-art across 15 distinct conditions. Code can be available at https://anonymous.4open.science/r/ALIEN/.

\end{abstract}

\section{Introduction}

Text-to-image diffusion models, such as Stable Diffusion \cite{Rombach_2022_CVPR} and DALL·E \cite{ramesh2022hierarchical}, have demonstrated impressive capabilities in generating high-quality images. To safeguard the intellectual property of generative models \cite{gowal2023identifying} and facilitate misuse tracking \cite{barrett2023identifying}, Governments are increasingly calling for regulations \cite{EU_AI_Act_2024,Biden_2023} to mandate watermark adoption. Robust and imperceptible watermarking of generated images has become a critical and urgent research focus.
Post-processing watermarking \cite{cox2007digital} is applied to the contents generated by the diffusion model, but its robustness is insufficient for reliable detection in real-world applications \cite{ren2024sok}. Some studies consider model distribution scenarios \cite{rezaei2024lawa,ci2024wmadapter,feng2024aqualora,wang2025sleepermark}. Diffusion models are fine-tuned to embed watermarks into the model parameters, which inevitably limits efficiency and scalability.

\begin{figure}[t]
  \begin{center}
    \centerline{\includegraphics[width=1\columnwidth]{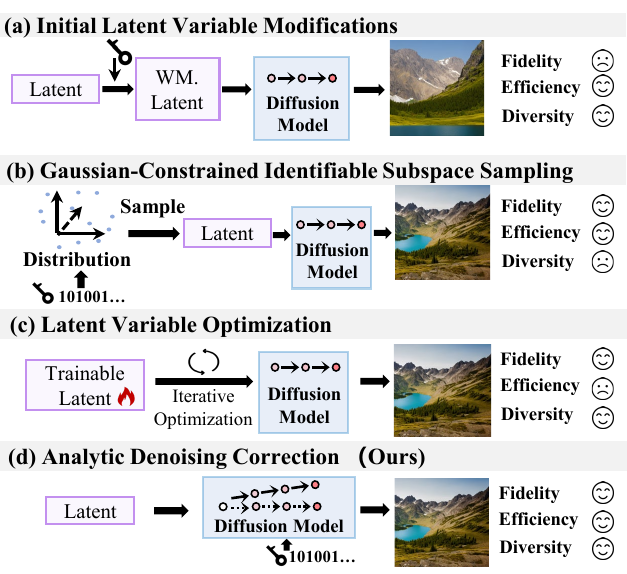}}
    \caption{(a) Latent modification, (b) Constrained sampling, (c) Iterative optimization, (d) Our ALIEN with principled embedding.}
    \label{intro_img}
    \vspace{-1.9em}
  \end{center}
\end{figure}

Recent research has concentrated on semantic watermarking \cite{lee2025semantic}, which aims to embed watermark signals into the semantic or latent features to better resist image-processing-based attacks \cite{zhao2024invisible}. Watermarking methods based on initial latent variable modifications \cite{wen2024tree,ci2025ringid} embed watermarks by directly modifying or adding perturbations to the initial latents (Fig.~\ref{intro_img}(a)). While the mechanism is intuitive, the mapping between the initial latent space and the final image is highly complex and nonlinear, making direct modifications prone to semantic drift. Furthermore, to maintain invisibility, the amount of modification to the initial latents is strictly limited, making it difficult to increase watermark capacity. 
To ensure high capacity and lossless watermark embedding, Watermarks based on the Gaussian-Constrained Identifiable Subspace Sampling \cite{yang2024gaussian,gunn2024undetectable} combined with cryptography, divide the initial latent space into non-overlapping identifiable subspaces, forcing users to sample from subspaces associated with specific watermark information (Fig.~\ref{intro_img}(b)). This rigid constraint on the initial latent space limits generative diversity. Watermarking methods based on optimization \cite{huang2024robin,zhang2024attack} perform watermark optimization in the latent space to better balance robustness and fidelity (Fig.~\ref{intro_img}(c)). However, their reliance on computationally intensive heuristic optimization to iteratively find the optimal watermark, leading to substantial training overhead and prone to local optima. Furthermore, detection of semantic watermarking typically depends on diffusion inversion, which limits applicability to reversible samplers \cite{lu2022dpm} and watermarks become undetectable when images are generated with irreversible samplers \cite{karras2022elucidating}. Attackers can exploit this vulnerability to remove watermarks by regenerating images with irreversible schedulers. \cite{an2024waves}. 

Despite the progress made by aforementioned semantic watermarking schemes in robust embedding, they circumvent a fundamental issue: the inability to derive a precise and efficient watermark embedding mechanism from the generative principles of diffusion models. Current optimization or constraint-based methods essentially avoid this problem, instead relying on computationally intensive optimization or sampling constraints. These approaches inherently limits the fidelity, efficiency, and diversity.


To address these issues, we propose an \underline{A}na\underline{l}ytical Watermark\underline{i}ng Framework for Controllabl\underline{e} Generatio\underline{n} (ALIEN). As shown in Fig.~\ref{intro_img}(d), unlike existing methods that rely on computationally intensive heuristic latent variable optimization or constraints on initial latent sampling, we start with the reverse process of the Stochastic Differential Equation of the diffusion model and present the first analytic derivation of the watermark residual propagation mechanism. 
Specifically, given a target watermark residual, we analytically derive a time-dependent modulation coefficient that transforms this target into a precise correction term for the noise prediction. By injecting this correction at each denoising step, we effectively modify the underlying probability flow of the diffusion model. This imposes a deterministic force that seamlessly guides the generation trajectory toward the watermarked state regardless of the sampling path, thereby eliminating the need for iterative optimization and ensuring compatibility with various samplers.
ALIEN achieves a principled watermark embedding pattern, rather than relying on heuristic methods. Without the need for iterative optimization, we inject the precisely modulated watermark into the noise prediction target at each step of the denoising process. 

Our contributions are three-fold: (1) We achieve the first analytic derivation of watermark propagation for seamless, low-inference-cost embedding, eliminating the need for iterative optimization. (2) Since the watermark is precisely compensated based on diffusion model, ALIEN preserves semantic consistency and visual fidelity by leveraging precise theoretical compensation for the watermark signal. (3) The ALIEN watermark embedding mechanism is independent of specific initial latent variables and sampler types (including irreversible samplers), solving the issue of strong reliance on reversible samplers of semantic watermarking.

\section{Related Work}
\textbf{Diffusion models} exhibit exceptional performance in image generation \cite{dhariwal2021diffusion}, leveraging the methodology \cite{ho2020denoising,song2020score} and the sampling techniques \cite{song2020denoising,song2020improved}. 
LDMs optimize image generation within the latent space of the pretrained Variational Autoencoder (VAE), which further accelerates the practical applications of diffusion models. 
During the inference phase, LDM first samples an initial latent $\mathbf{z}_T \in \mathbb{R}^{c \times w \times h}$ from a standard Gaussian distribution $\mathcal{N}(\mathbf{0}, \mathbf{I})$, where $T$ denotes the total time step of the diffusion model. Following iterative denoising, the latent vector evolves into the noise-free representation $\mathbf{z}_0$. The final image $\mathbf{x}_0$ is  reconstructed by the VAE decoder from $\mathbf{z}_0$. Safeguarding the intellectual property of generated content and preventing misuse have become critical research priorities. This paper focuses on critical issues of fidelity, efficiency, and controllability in existing approaches for integrating watermarking into the diffusion process.

\begin{figure*}[t]
\centering
\includegraphics[width=1.9\columnwidth]{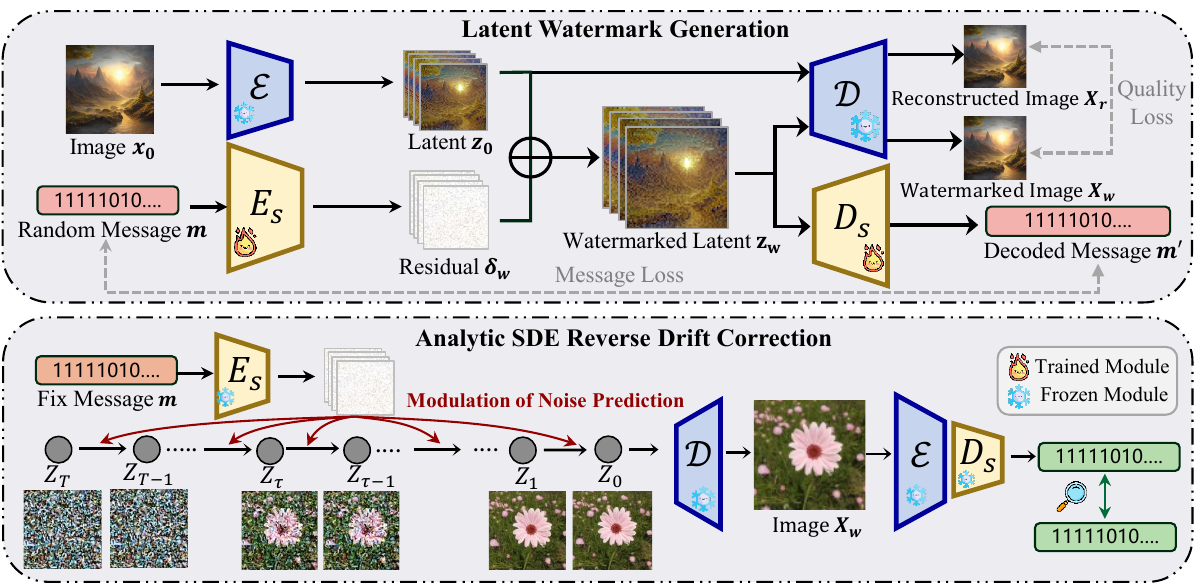} %
\caption{
The ALIEN framework consists of two main stages: (Top) Imperceptible Latent Watermark Generation, where a secret encoder $E_s$ and decoder $D_s$ are trained to embed a message $m$ into a robust latent residual $\delta_w$ while preserving image quality. (Bottom) Analytic SDE Reverse Drift Correction, which applies the time-dependent modulation to the noise prediction. This principled correction steers the generative trajectory to satisfy the watermark constraint in a sampler-agnostic manner.}
\label{Framework}
\end{figure*}

\textbf{Watermarking in Latent Diffusion Models} aims at tracking the origin and ensuring the accountability of the generated content. 
For a comprehensive context regarding the taxonomy of existing watermarking methods
a detailed review is provided in Appendix \ref{app:related}. 
We specifically focus on prior work integrating watermarking during the diffusion process. Latent modification Watermark such as Tree-Ring \cite{wen2024tree} and RingID \cite{ci2025ringid} embed watermarks in the Fourier space of the initial latent which leads to semantic drift. Constrained sampling watermark such as Gaussian Shading \cite{yang2024gaussian}, GaussMarker \cite{li2025gaussmarker} and PRC \cite{gunn2024undetectable} utilize cryptographic principles to modify the sampling pattern, but strict constraints limit the sampling range, resulting in reduced controllability. Optimization-based methods such as Zodiac \cite{zhang2024attack} and Robin \cite{huang2024robin}, optimize latent variables for higher semantic consistency, but reliance on heuristic optimization incurs computational costs. The detection of aforementioned methods depends on diffusion inversion, limiting their function to reversible samplers and is ineffective with irreversible samplers. Our method achieves low-overhead and universally applicable watermark via analytic derivation.

\section{Methodology}






\subsection{Framework of ALIEN}
We extend to achieve applicability and theoretical controllability of semantic watermarking. As demonstrated in Fig.~\ref{Framework}, our ALIEN embeds the watermark in intermediate diffusion states to manipulate the generation trajectory through precise correction of the probability flow. 
To generate imperceptible watermark residuals, we implement an Imperceptible Latent Watermark Generation module to generate the invisible and robust watermark residual in the latent space. 
For principled embedding, we propose Analytic SDE Reverse Drift Correction, which analytically derives the necessary modifications to the probability flow required by the diffusion model for watermark embedding under Variance Preserving Stochastic Differential Equation (VP-SDE) \cite{song2020score}, providing an explicit correction target for noise prediction that is compatible with both stochastic and deterministic sampling processes. This derived target is implemented via the Modulation of Noise Prediction Target module, which specifies the requisite adjustment to the model's noise prediction output, thereby realizing the controlled and imperceptible semantic watermarking.

\subsection{Imperceptible Latent Watermark Generation}
To generate an imperceptible yet robust watermark residual, we jointly train a secret encoder $E_s$ and a watermark decoder $D_s$. Ideally, the watermarked latent representation $z_w$ should be conditioned on both the input latent $z_0$ and the message $m$ to enhance imperceptibility. However, utilizing a $z_0$-dependent watermark becomes impractical as the intermediate latent states of the diffusion model are not readily accessible for deterministic conditioning. Therefore, we opt to embed a cover-agnostic watermark prominent in the latent space. Specifically, the secret residual $\delta_w = E(m)$ is added to the input latent $z_{0}$, forming the watermarked latent $z_w = z_{0} + \delta_w$. The watermarked image $x_w$ is generated as $x_w = \mathcal{D}(z_w)$, and the message is extracted by applying the decoder $D_s$ to $z_w$, yielding $m' = D_s(z_w)$. We employ the Binary Cross-Entropy loss to optimize for the accuracy of message extraction between the original message $m$ and the decoded message $m'$.

\begin{figure}[t]
  \begin{center}
\centerline{\includegraphics[width=\columnwidth]{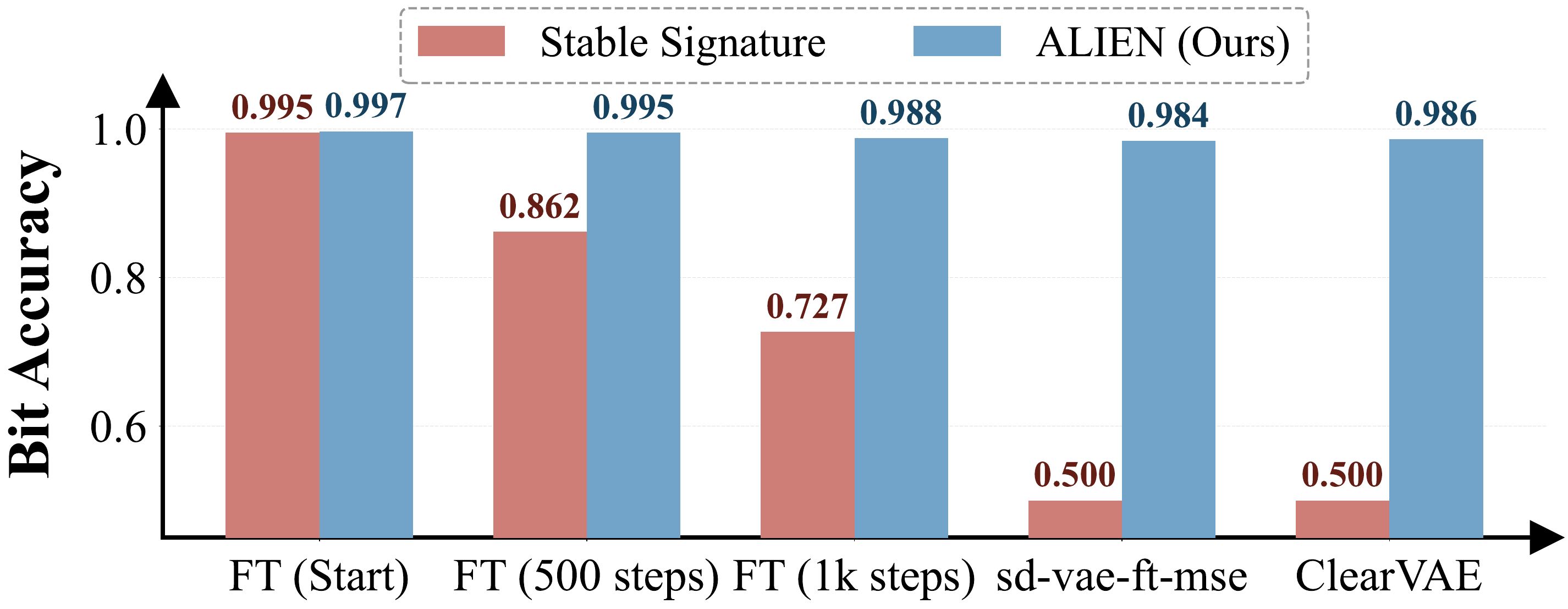}}
    \caption{Bit accuracy comparison against Stable Signature across VAE fine-tuning and replacement.}
    \label{vae replace}
  \end{center}
  \vspace{-0.8em}
\end{figure}

To ensure the visual consistency of the watermark, we compute the LPIPS loss \cite{zhang2018unreasonable} and the Mean Squared Error loss between the watermarked image $x_w$ and the reconstructed image $x_r$, rather than between $x_w$ and the original image $x_o$. This choice is necessary because the VAE compression and reconstruction already introduce a measurable irrecoverable quality gap between $x_o$ and $x_r$. Optimizing against $x_o$ would necessitate the watermark training process to compensate for VAE reconstruction errors, which would increase complexity and hinder embedding effectiveness. Our training objective can be summarized below, where $\lambda_1$ and $\lambda_2$ are coefficients:

\begin{equation}\label{eq:ltotoal}
\mathcal{L}_{T} = \mathcal{L}_{BCE}^{(m,m')} + \lambda_1\mathcal{L}_{LPIPS}^{(x_r,x_w)} + \lambda_2\mathcal{L}_{MSE}^{(x_r,x_w)},
\end{equation}
where $\mathcal{L}_{BCE}$ is the Binary Cross-Entropy loss, $\mathcal{L}_{LPIPS}$ is the Learned Perceptual Image Patch Similarity loss \cite{zhang2018unreasonable}, and $\mathcal{L}_{MSE}$ is the Mean Squared Error loss.
Existing studies \cite{wang2025sleepermark,meng2024latent} demonstrate that injecting and detecting watermarks in the latent space can inherently resist various common distortions. We follow prior practice by omitting the distortion layer during training (validated in Tab.~\ref{tab:robustness_final_with_avg}). The effectiveness of watermark remains unaffected even if an adversary fine-tunes U-Net $\theta$ and $\mathcal{D}$ on clean images and uses a fine-tuned latent decoder $\mathcal{D}'$ to generate images (validated in Fig.~\ref{vae replace}).

\newcommand{\zinit}{\mathbf{z}_T}
\newcommand{\zstep}{\mathbf{z}_t}
\newcommand{\zprev}{\mathbf{z}_{t-1}}
\newcommand{\zfinal}{\mathbf{z}_0}
\newcommand{\ximg}{\mathbf{x}}
\newcommand{\wmres}{\mathbf{\delta}_{wm}}
\newcommand{\secret}{\mathbf{m}}
\newcommand{\Es}{E_s}
\newcommand{\Ds}{D_s}
\newcommand{\Unet}{\theta}                 
\newcommand{\epred}{\mathbf{\epsilon}_{\theta}^t} 
\newcommand{\Correction}{\Delta \mathbf{\epsilon}}
\newcommand{\GammaT}{\gamma_t}
\newcommand{\alphaBar}{\bar{\alpha}_t}
\newcommand{\DVAE}{\mathcal{D}_{VAE}}
\newcommand{\EVAE}{\mathcal{E}_{VAE}}

\begin{algorithm}[t]
\caption{ALIEN Watermarking}
\label{alg:alien_compact}
\begin{algorithmic}[1]

\STATE \textcolor{gray}{\hrulefill\ \textsc{Phase I: Embedding}\ \hrulefill}

\STATE \textbf{Input}: Pre-trained U-Net $\Unet$, Encoder $\Es$, Scheduler $S$, VAE $\DVAE$, Prompt $\mathbf{c}$, Secret $\secret$, Strength $\lambda$, Interval $[T_{start}, T_{end}]$
\STATE \textbf{Output}: Watermarked Image $\ximg_{wm}$

\STATE $\wmres \leftarrow \Es(\secret)$
\STATE $\zstep \sim \mathcal{N}(\mathbf{0}, \mathbf{I})$

\STATE \textbf{for} $t = T$ \textbf{down to} $1$ \textbf{do}
    \STATE \quad $\epred \leftarrow \text{CFG}(\Unet, \zstep, S.\text{steps}[t], \mathbf{c})$

    \STATE \quad \textbf{if} $T_{start} \le t \le T_{end}$ \textbf{then}
        \STATE \quad \quad $\epred \leftarrow \epred - \lambda \cdot \left(\frac{\sqrt{\bar{\alpha}_t}}{\sqrt{1 - \bar{\alpha}_t}}\right) \cdot \wmres$ 
    \STATE \quad \textbf{end if}
    
    \STATE \quad $\zstep \leftarrow S.\text{step}(\epred, t, \zstep).\text{prev\_sample}$
\STATE \textbf{end for}
\STATE $\ximg_{wm} \leftarrow \DVAE(\zstep)$

\STATE \textcolor{gray}{\hrulefill\ \textsc{Phase II: Extraction}\ \hrulefill}

\STATE \textbf{Input}: Image $\ximg_{wm}$, $\EVAE$, Decoder $\Ds$ 
\STATE \textbf{Output}: Secret $\secret^{\prime}$

\STATE \textbf{Return} $\Ds(\EVAE(\ximg_{wm}))$ 

\end{algorithmic}
\end{algorithm}
\begin{figure}[t]
  \centering
  \captionsetup[subfloat]{skip=4pt} 
  
  \subfloat[Noise Prediction Norm Discrepancy]{
    \includegraphics[width=1\linewidth]{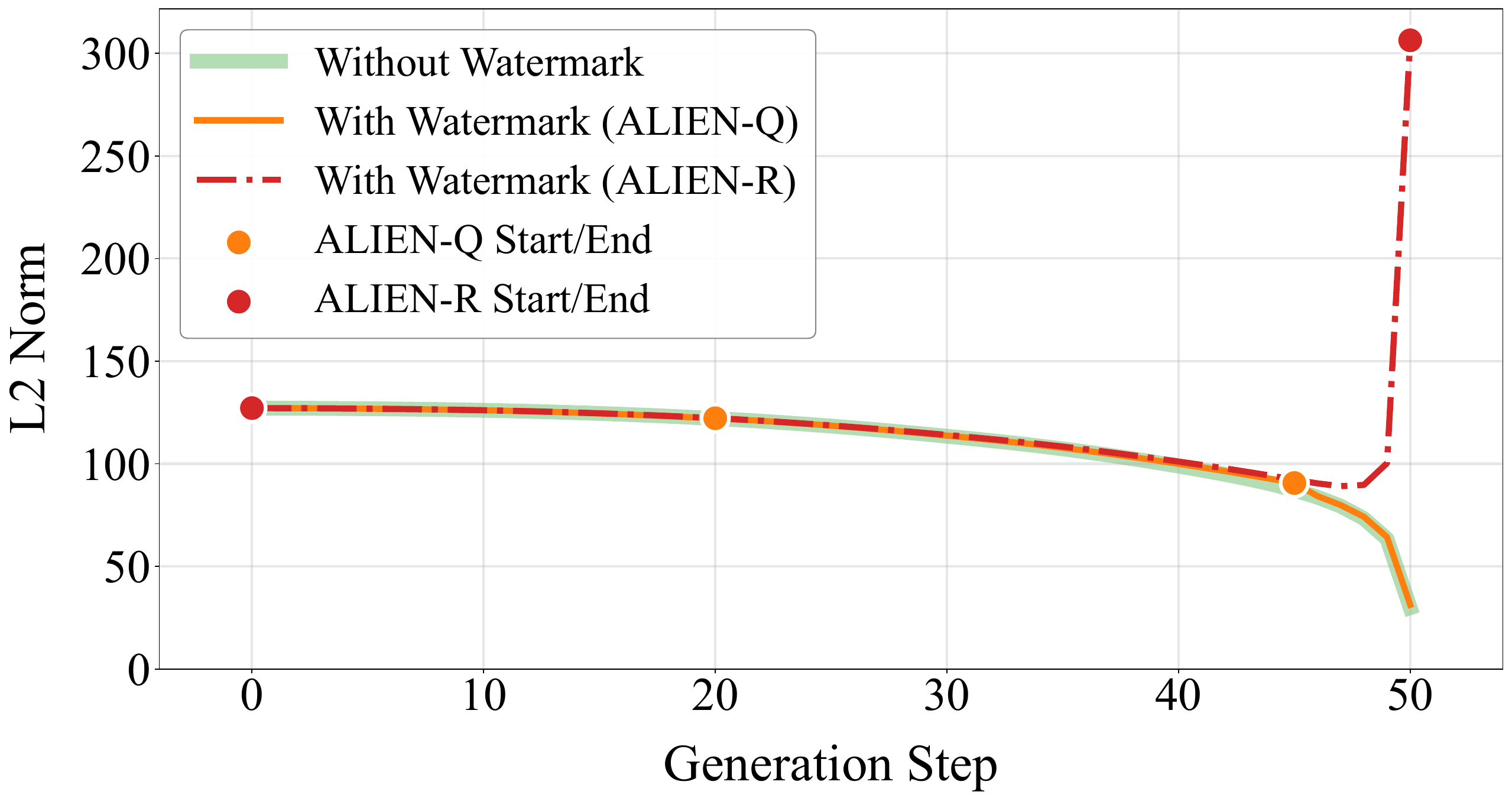}
    \label{fig:noise_norm}
  } \hfill
  \subfloat[Injection Strength Schedule]{
    \includegraphics[width=0.95\linewidth]{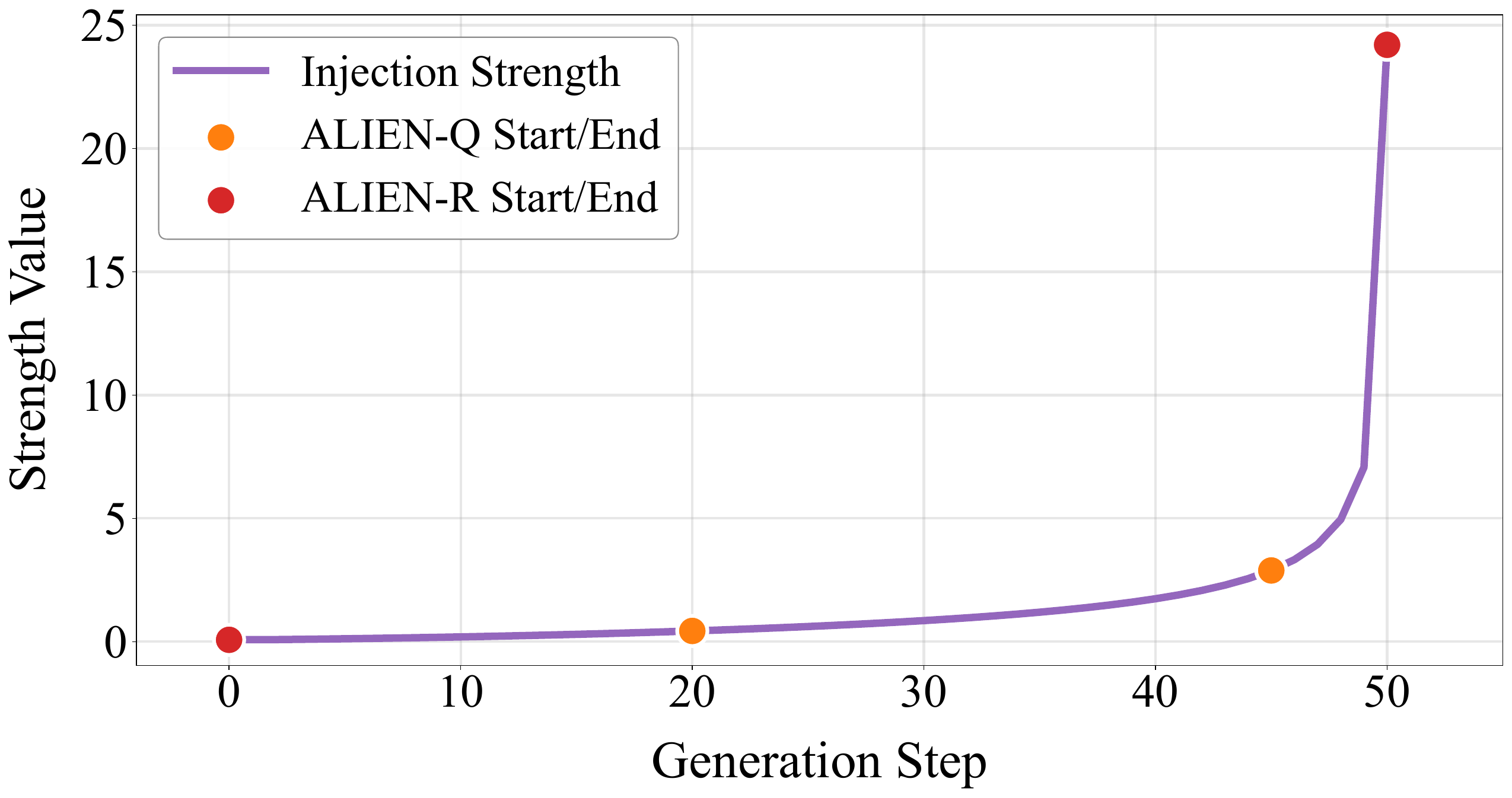}
    \label{fig:injection_strength}
  }
  \caption{\textbf{Comparison of ALIEN-R and ALIEN-Q.} 
  (a) The L2 norm of noise prediction during the diffusion process. 
  (b) The evolution of injection strength $\frac{\sqrt{\bar{\alpha}_t}}{\sqrt{1 - \bar{\alpha}_t}}$.}
  \label{fig:comparison}
\end{figure}
\subsection{Analytic SDE Reverse Drift Correction}

\begin{figure*}[t]
\centering
\includegraphics[width=2\columnwidth]{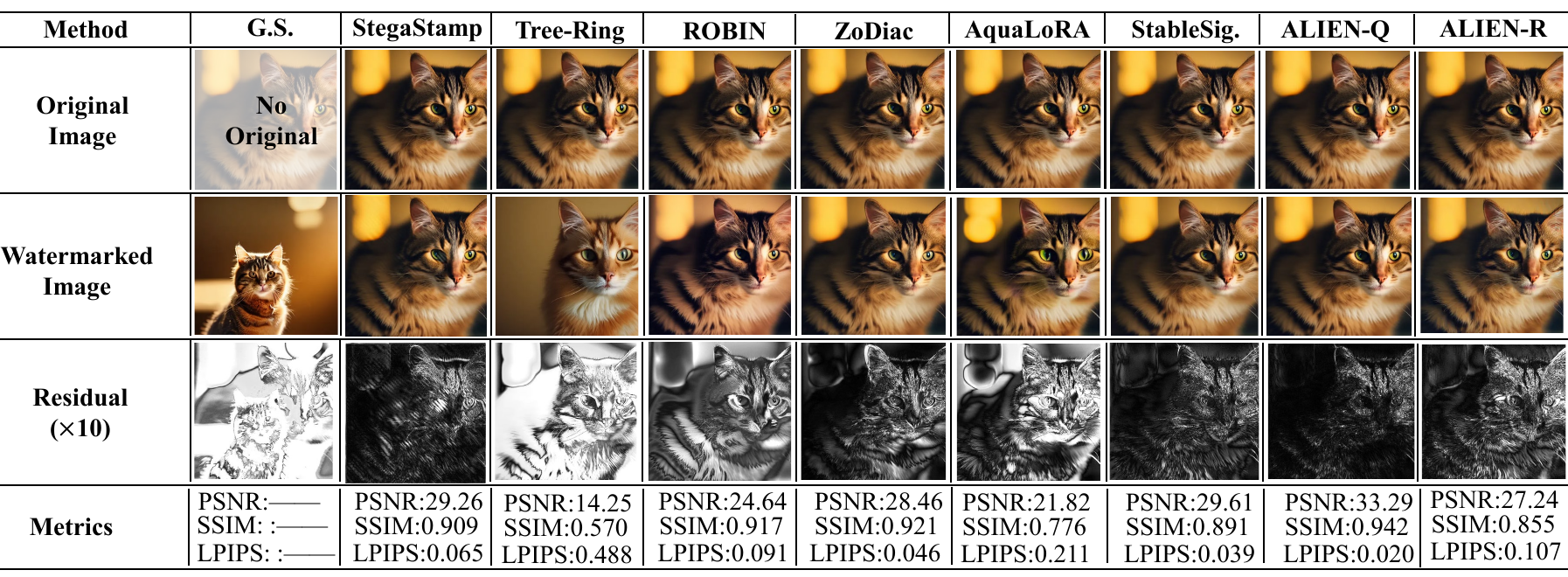} %
\caption{Qualitative comparison of watermarked samples and $10\times$ magnified residuals. We compare ALIEN with baselines covering post-processing (StegaStamp), latent modification (Tree-Ring), optimization (ROBIN, ZoDiac), and fine-tuning (AquaLoRA, StableSig.).}
\label{Fidelity Comparison}
\end{figure*}
To achieve precise watermark embedding over the generative process, we leverage the VP-SDE to analytically derive the exact probability flow correction required for watermark embedding.

\textbf{$\mathbf{z}_0$ Constraint.} Our goal is to embed a predetermined watermark residual $\mathbf{\delta}_{wm}$ into the final denoised latent $\mathbf{z}_0$, resulting in the $\mathbf{z}_0$-space constraint  $\hat{\mathbf{z}}_0^{wm} = \hat{\mathbf{z}}_0^{orig} + \mathbf{\delta}_{wm}$,
where $\hat{\mathbf{z}}_0^{wm}$ and $\hat{\mathbf{z}}_0^{orig}$ are the clean data estimates predicted by the U-Net for the current latent state $\mathbf{z}_t$, with and without the watermark, respectively. 

\textbf{Score Function.} Under the VP-SDE, The diffusion process is defined by a forward SDE that gradually perturbs clean data into noise:

$$
\mathrm{d}\mathbf{z} = \mathbf{f}(\mathbf{z}, t)\mathrm{d}t + g(t)\mathrm{d}\mathbf{w},
$$

where $\mathbf{f}(\mathbf{z}, t): \mathbb{R}^d \to \mathbb{R}^d$ denotes the drift coefficient, while $g(t) \in \mathbb{R}$ represents the scalar diffusion coefficient, with $\mathbf{w}(t)$ denoting the standard Wiener process.
The reverse generation process of the diffusion model is governed by a reverse-time stochastic differential equation:
\begin{equation}
    \mathrm{d}\mathbf{z} = \underbrace{[\mathbf{f}(\mathbf{z}, t) - g^2(t) \nabla_{\mathbf{z}} \log p_t(\mathbf{z})]}_{\text{Reverse Drift } \mathbf{F}_{rev}} \mathrm{d}t + g(t) \mathrm{d}\bar{\mathbf{w}},
\end{equation}
where $\bar{\mathbf{w}}$ denotes the standard Wiener process in reverse time.
To realize the watermark $z_0$ constraint, we precisely control the probability flow, which requires determining the necessary correction $\Delta \mathbf{F}_{rev}$ for the reverse SDE drift term. Since $\Delta \mathbf{F}_{rev}$ relies on the score function $\nabla_{\mathbf{z}_t} \log p_t(\mathbf{z}_t)$, analytically quantifying the difference in the score function becomes a direct path to derive $\Delta \mathbf{F}_{rev}$. The score function is exactly and linearly related to the difference between $\mathbf{z}_t$ and its denoised estimate $\hat{\mathbf{z}}_0$.
This analytical relationship is mathematically described as follows:
\begin{equation}
    \nabla_{\mathbf{z}_t} \log p_t(\mathbf{z}_t) = -\frac{1}{1 - \bar{\alpha}_t} (\mathbf{z}_t - \sqrt{\bar{\alpha}_t} \hat{\mathbf{z}}_0).
    \label{eq:analyze}
\end{equation}

Applying the analytical mapping Eq.\ref{eq:analyze} allows us to analytically determine the difference in the score function induced by the $\mathbf{z}_0$ constraint, thereby translating the watermark residual $\mathbf{\delta}_{wm}$ into the required shift in the score function space.
\begin{table}[t]
    \centering
    \small
    \caption{Quantitative Comparison of Visual Quality and Fidelity across Watermarking Schemes. D.S. denotes the Dreamsim \cite{fu2023dreamsim} metric.}
    \setlength{\tabcolsep}{5pt}
    \label{tab:fidelity_comparison_no_bold}
    \begin{tabular}{lcccccc}
        \toprule
         Method & \footnotesize{FID} & \footnotesize{CLIP} & \footnotesize{PSNR} & \footnotesize{SSIM} &\footnotesize{SIFID} & \footnotesize{D.S.}\\
        \midrule
        No WM & 24.31 & 0.3368 & --- & --- & --- & --- \\
        StegaS. & 24.56 & 0.3363 & 28.59 & 0.878 & 0.189& 0.021 \\
        ZoDiac & --- & --- & 28.01 & 0.922 & 0.121& 0.022 \\
        AquaL. & 24.79 & 0.3366 & 17.07 & 0.664 & 0.183& 0.139 \\
        TreeR. & 24.63 & 0.3370 & 12.76 & 0.429 & 0.741& 0.291 \\
        G.S. & 24.42 & \cellcolor{gray!15}0.3378 & --- & --- & --- & --- \\
        ROBIN & 24.61 & 0.3366 & 22.96 & 0.756 & 0.212& 0.057 \\
        StableS. & 24.56 & 0.3367 & 29.09 & 0.878 & 0.105& 0.011 \\
        \midrule
        \cellcolor{gray!15}ALIEN-Q & \cellcolor{gray!15}24.29 & {0.3369} & \cellcolor{gray!15}32.41 & \cellcolor{gray!15}0.949 & \cellcolor{gray!15}0.023 & \cellcolor{gray!15}0.003\\
        ALIEN-R & 24.74 & 0.3366 & 20.42 & 0.745 & 0.227 & 0.061\\
        \bottomrule
    \end{tabular}
\end{table}



\begin{table}[t]
    \centering
    \caption{Stability Evaluation measured by Detection Confidence. We assess robustness across Samplers (DPM++ SDE, Euler a., DPM2 a.), Inference Steps ($25, 50$), Guidance Scale ($10, 20$), and Model Versions (v1.5, v2.1).}
    \label{tab:stability_bit_acc}
    
    \resizebox{\columnwidth}{!}{%
    \setlength{\tabcolsep}{3pt} 
    \begin{tabular}{l ccc cc cc cc}
        \toprule
        \multirow{2}{*}{\textbf{Method}} 
        & \multicolumn{3}{c}{\textbf{Sampler}} 
        & \multicolumn{2}{c}{\textbf{Steps}} 
        & \multicolumn{2}{c}{\textbf{Scale}} 
        & \multicolumn{2}{c}{\textbf{Model}} \\
        
        \cmidrule(lr){2-4} \cmidrule(lr){5-6} \cmidrule(lr){7-8} \cmidrule(lr){9-10}
        
        & DPM-SDE & Eulera & DPM2a 
        & 25 & 50 
        & 10 & 20 
        & v1.5 & v2.1 \\
        \midrule
        
        \rowcolor{gray!15}StegaS.  & 0.999 & 0.999 & 0.999 & 0.999 & 0.999 & 0.999 & 0.999 & 0.990 & 0.999 \\
        \rowcolor{gray!15}StableS. & 0.999 & 0.999 & 0.999 & 0.999 & 0.999 & 0.999 & 0.999 & 0.999 & 0.999 \\
        TreeR.   & 0.000    & 0.000    & 0.000    & 0.963    & 0.963    & 0.963    & 0.963    & 0.958    & 0.963    \\ 
        G.S.     & 0.557 & 0.586 & 0.552 & \cellcolor{gray!15}0.999 & \cellcolor{gray!15}0.999 & \cellcolor{gray!15}0.999 & \cellcolor{gray!15} 0.999 & \cellcolor{gray!15}0.999 & \cellcolor{gray!15}0.999 \\ 
        AquaL.   & 0.952 & 0.953 & 0.939 & 0.945 & 0.955 & 0.940 & 0.939 & 0.954 & \nna \\ 
        \midrule 
        
        ALIEN-Q  & 0.989 & 0.979 & 0.972 & 0.985 & 0.990 & 0.982 & 0.975 & 0.989 & 0.991 \\ 
        \rowcolor{gray!15} 
        ALIEN-R  & 0.999 & 0.999 & 0.999 & 0.999 & 0.999 & 0.999 & 0.999 & 0.999 & 0.999 \\ 
        \bottomrule
    \end{tabular}%
    }
\end{table}


\newcommand{\graycell}[1]{\setlength{\fboxsep}{0pt}\colorbox{gray!20}{#1}}

\begin{table*}[t]
    \centering
    \caption{\textbf{Robustness Evaluation across variant Regenerative Attacks.} 
    Each cell reports performance under three attacks: \textbf{Regeneration $\to$ Rinse-2X $\to$ Rinse-4X}. 
    Values denote \textbf{Detection Confidence/ TPR@1\%FPR}. }
    \label{tab:robustness_by_scheduler}
    
    \setlength{\tabcolsep}{2pt}
    \renewcommand{\arraystretch}{1.35} 
    \scriptsize 
    
    \resizebox{\textwidth}{!}{
        \begin{tabular}{l | c | c | c | c} 
            \toprule
            \multirow{2}{*}{\textbf{Method}} & \multicolumn{4}{c}{\textbf{Scheduler (Cell Format: Regen. Data $|$ Rinse-2X Data $|$ Rinse-4X Data)}} \\
            \cmidrule(lr){2-5}
            
             & \textbf{DDIM} & \textbf{DPM++ 2M SDE} & \textbf{Euler a} & \textbf{DPM2 a} \\
            \midrule
            
            StegaS.  
            & 0.692/0.957 $|$ 0.578/0.347 $|$ 0.531/0.102 
            & 0.636/0.575 $|$ 0.562/0.182 $|$ 0.502/0.000 
            & 0.624/0.631 $|$ 0.542/0.180 $|$ 0.513/0.000 
            & 0.615/0.591 $|$ 0.556/0.120 $|$ 0.495/0.000 \\
            
            ZoDiac   
            & 0.938/1.000 $|$ 0.743/0.104 $|$ 0.527/0.032 
            & 0.693/0.585 $|$ 0.670/0.431 $|$ 0.562/0.125 
            & 0.643/0.315 $|$ 0.643/0.250 $|$ 0.628/0.210 
            & 0.833/0.650 $|$ 0.572/0.281 $|$ 0.520/0.152 \\
            
            TreeR.   
            & 0.868/1.000 $|$ 0.724/0.861 $|$ 0.408/0.489 
            & 0.855/0.918 $|$ 0.824/0.905 $|$ 0.753/0.851 
            & 0.821/0.941 $|$ 0.449/0.512 $|$ 0.277/0.306 
            & 0.647/0.755 $|$ 0.531/0.656 $|$ 0.326/0.427 \\
            
            G.S.     
            & 0.978/1.000 $|$ 0.898/1.000 $|$ 0.845/1.000 
            & 0.943/1.000 $|$ 0.953/1.000 $|$ 0.926/1.000
            & 0.946/1.000 $|$ 0.847/1.000 $|$ 0.695/0.905 
            & 0.923/1.000 $|$ 0.858/1.000 $|$ 0.711/0.915 \\
            
            AquaL.   
            & 0.757/0.742 $|$ 0.627/0.145 $|$ 0.573/0.000 
            & 0.879/0.942 $|$ 0.833/0.858 $|$ 0.696/0.486 
            & 0.663/0.225 $|$ 0.618/0.078 $|$ 0.551/0.000 
            & 0.679/0.371 $|$ 0.604/0.086 $|$ 0.536/0.029 \\
            
            ROBIN    
            & \nna/0.638 $|$ \nna/0.242 $|$ \nna/0.153 
            & \nna/1.000 $|$ \nna/0.728 $|$ \nna/0.275 
            & \nna/0.980 $|$ \nna/0.705 $|$ \nna/0.486 
            & \nna/0.075 $|$ \nna/0.105 $|$ \nna/0.100 \\
            
            StableS. 
            & 0.496/0.000 $|$ 0.477/0.000 $|$ 0.521/0.000 
            & 0.512/0.000 $|$ 0.512/0.000 $|$ 0.510/0.000 
            & 0.498/0.000 $|$ 0.520/0.000 $|$ 0.513/0.000 
            & 0.506/0.000 $|$ 0.485/0.000 $|$ 0.502/0.000 \\
            
            \midrule
            
            ALIEN-Q 
            & 0.848/1.000 $|$ 0.752/0.865 $|$ 0.618/0.342 
            & 0.935/1.000 $|$ 0.898/1.000 $|$ 0.842/0.942 
            & 0.866/1.000 $|$ 0.661/0.581 $|$ 0.563/0.163 
            & 0.857/1.000 $|$ 0.673/0.636 $|$ 0.581/0.124 \\
            
            ALIEN-R
            & 0.999/1.000 $|$ 0.908/1.000 $|$ 0.829/1.000
            & 0.999/1.000 $|$ 0.989/1.000 $|$ 0.967/1.000 
            & 0.988/1.000 $|$ 0.908/1.000 $|$ 0.731/0.727 
            & 0.989/1.000 $|$ 0.878/1.000 $|$ 0.742/0.875 \\
            \bottomrule
        \end{tabular}
    }
\end{table*}
\textbf{Probability Flow Correction.} Based on the analytical structure of VP-SDE for the reverse drift term $\mathbf{F}_{rev}$, we derive the required reverse SDE drift correction $\Delta \mathbf{F}_{rev}$ to enforce the watermark constraint (Details provided in Appendix \ref{app:theory}). The resulting correction term $\Delta \mathbf{F}_{rev}$ is analytically defined by the following expression:
\begin{equation}
    \Delta \mathbf{F}_{rev}(\mathbf{z}_t, t) = - g^2(t) \frac{\sqrt{\bar{\alpha}_t}}{1 - \bar{\alpha}_t} \mathbf{\delta}_{wm}.
    \label{eq:rev}
\end{equation}
This analytical derivation Eq.\ref{eq:rev} provides the theoretical foundation for the ALIEN framework. It precisely quantifies the correction $\Delta \mathbf{F}_{rev}$ required by the watermark objective $\mathbf{\delta}_{wm}$ within the SDE probability flow framework. Since this correction targets the score function of the latent distribution, it demonstrates that the watermark is sampler-agnostic.



\subsection{Modulation of Noise Prediction Target}
To achieve the practical implementation of the SDE drift correction $\Delta \mathbf{F}_{rev}$, we analytically relate the SDE reverse drift correction to the noise prediction offset of U-Net within the VP-SDE framework (Details provided in Appendix \ref{app:theory}), we derive the required compensation signal $\Delta \mathbf{\epsilon}_{wm}$ necessary to enforce the $\mathbf{z}_0$ constraint:

\begin{equation}
\Delta \mathbf{\epsilon}_{wm} = - \frac{\sqrt{\bar{\alpha}_t}}{\sqrt{1 - \bar{\alpha}_t}} \mathbf{\delta}_{wm}.
\label{eq:pred}
\end{equation}

Refer to Eq.~(\ref{eq:pred}), to realize a fixed watermark residual $\mathbf{\delta}_{wm}$ in the $\mathbf{z}_0$ space, the U-Net will introduce a time-dependent compensation signal $\Delta \mathbf{\epsilon}_{wm}$ proportional to the watermark residual in its noise prediction. As illustrated in Fig.~\ref{fig:comparison}, the injection coefficient exhibits a surge approaching $\mathbf{z}_0$, implying that late-stage injection yields stronger constraints but risks visual fidelity. We adjust the injection magnitude and timestep range to formulate two configurations: ALIEN-Q (Quality-Oriented) and ALIEN-R (Robustness-Oriented). The procedure is outlined in Alg.~\ref{alg:alien_compact}.


\section{Experiments}
\subsection{Experimental Setup}
\noindent\textbf{Watermarking Baselines.} We compare ALIEN with mainstream latent semantic watermarking schemes across three primary categories: the initial latent variable modification method Tree-Ring \cite{wen2024tree}; the constrained sampling method Gaussian Shading \cite{yang2024gaussian}; and the iterative optimization methods ROBIN \cite{huang2024robin} and Zodiac \cite{zhang2024attack}. Additionally, we select the fine-tuning based method Stable Signature \cite{fernandez2023stable} and Aqualora \cite{feng2024aqualora}, as well as the post-processing image watermarking technique StegaStamp \cite{tancik2020stegastamp}. Implementation details can be found in Appendix \ref{app:training_details}.


\begin{figure}[t]
\centering
\includegraphics[width=0.95\columnwidth]{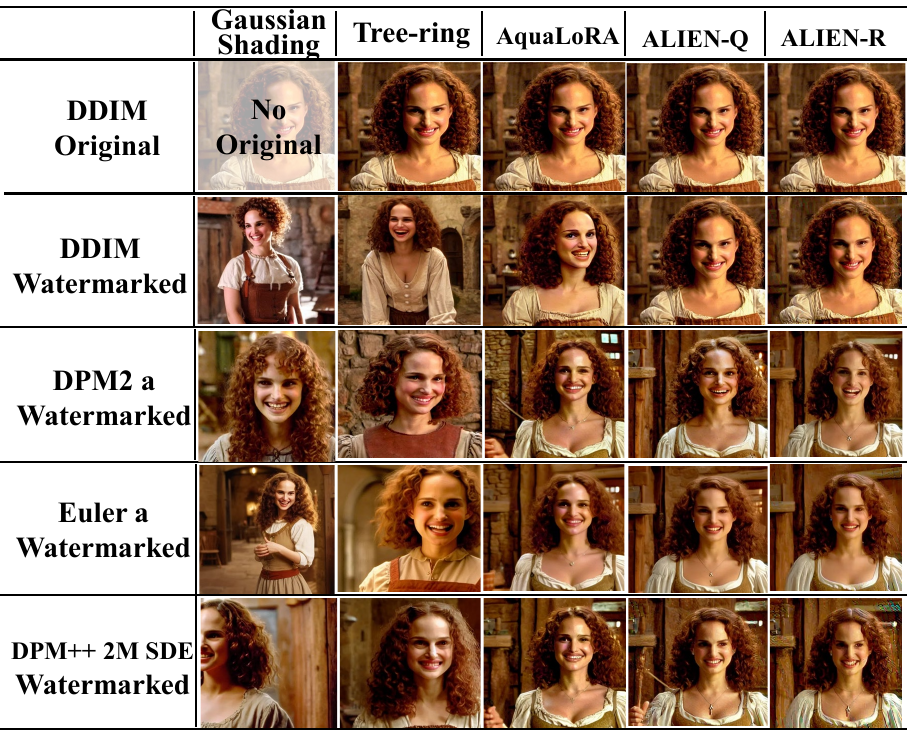} %
\caption{Visual Comparison under Different Generation Schedulers (DDIM, DPM2 a, Euler a, and DPM++ 2M SDE).}
\label{irreverse_scheme}
\vspace{-0.7em}
\end{figure}

\textbf{Models and Tasks.} Following existing research \cite{yang2024gaussian}, we implement the ALIEN framework on Stable Diffusion v1.5 and Stable Diffusion v2.1. 
We evaluate our method across two tasks:
(1) \textbf{Detection.} All compared methods are configured as single-bit watermarks with a unified pattern. 
The detection threshold is set individually for each method to achieve a False Positive Rate (FPR) of approximately 1\%. 
(2) \textbf{Traceability.} Multi-bit watermarking methods are evaluated using Bit Accuracy. Single-bit methods, including Tree-ring, Zodiac, and Robin, are excluded due to their lack of watermark capacity. We use prompts from Stable Diffusion-Prompts (SDP) \cite{Gustavosta2023}, setting the Guidance Scale to 7.5 and the number of sampling steps to 50. Both TPR@1\%FPR and Bit Accuracy are computed over 350 watermarked images.

\subsection{Comparison to Baselines}
\textbf{Watermark Fidelity.} We employ PSNR \cite{hore2010image} and SSIM \cite{wang2004image} for pixel fidelity, LPIPS \cite{zhang2018unreasonable} and DreamSim \cite{fu2023dreamsim} for perceptual similarity, FID \cite{heusel2017gans} and SIFID \cite{yang2024sifid} for realism, and CLIP Score \cite{Radford2021LearningTV} for semantic alignment. As shown in Tab.~\ref{tab:fidelity_comparison_no_bold}, ALIEN-Q achieves state-of-the-art imperceptibility, significantly outperforming all baselines. It yields the highest pixel fidelity of 32.41 dB PSNR and the best perceptual quality of 0.003 DreamSim, while maintaining FID and CLIP scores comparable to non-watermarked images.

\begin{figure}[t]
\centering
\includegraphics[width=1\columnwidth]{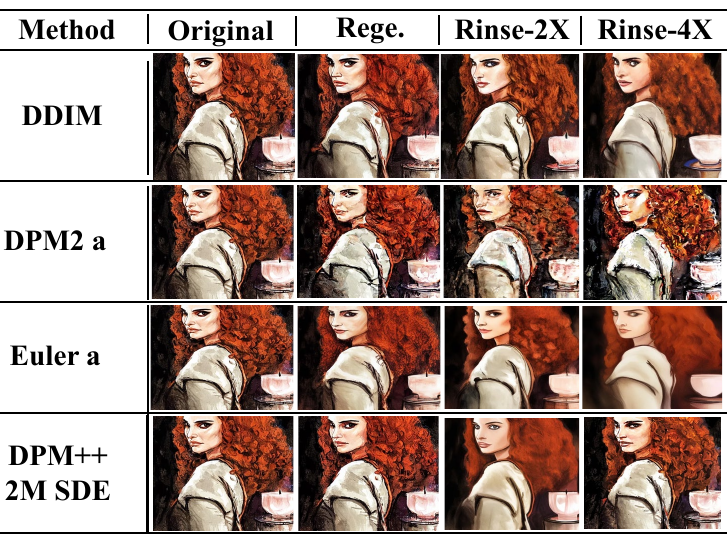} %
\caption{Visual Comparison under Regeneration Attacks (DDIM, DPM2 a, Euler a, and DPM++ 2M SDE).}
\label{diff_atk}
\vspace{-0.4em}
\end{figure}

\begin{table*}[htbp]
    \centering
    \caption{Comprehensive Comparison of Watermark Robustness. Performance is reported in decimal scale (0-1). Higher values are better ($\uparrow$). The \textbf{Avg.} column represents the mean performance across all 11 attacks (excluding No Attack). Top section: TPR@1\%FPR; Bottom section: Bit Accuracy.}
    \label{tab:robustness_final_with_avg}
    \setlength{\tabcolsep}{4.0pt} 
    \footnotesize 
    
    \begin{tabular}{cc c cc ccc ccc ccc c} 
        \toprule
        \multirow{2}{*}{Metrics} & \multirow{2}{*}{Method} & \multirow{2}{*}{\shortstack{No\\Attack}} & \multicolumn{2}{c}{Photometric} & \multicolumn{3}{c}{Degradation} & \multicolumn{3}{c}{Geometric} & \multicolumn{3}{c}{Generative} & \multirow{2}{*}{\textbf{Avg.}} \\
        
        \cmidrule(lr){4-5} \cmidrule(lr){6-8} \cmidrule(lr){9-11} \cmidrule(lr){12-14} 
        
        & & & Bright. & Contr. & JPEG & Blur & Noise & ReScale & C.C. & R.C. & VAE-B & VAE-C & Diff. & \\
        \midrule 
        
        \multirow{10}{*}{\centering \shortstack{TPR@\\Thre1\%FPR}} 
        & StegaS. & \cellcolor{gray!15}1.000 & 0.964 & \cellcolor{gray!15}1.000 & \cellcolor{gray!15}1.000 & \cellcolor{gray!15}1.000 & \cellcolor{gray!15}1.000 & \cellcolor{gray!15}1.000 & 0.964 & 0.975 & \cellcolor{gray!15}1.000 & \cellcolor{gray!15}1.000 & \cellcolor{gray!15}1.000 & 0.991 \\ 
        & StableS. & \cellcolor{gray!15}1.000 & 0.958 & \cellcolor{gray!15}1.000 & 0.781 & 0.973 & 0.042 & 0.935 & \cellcolor{gray!15}1.000 & \cellcolor{gray!15}1.000 & 0.000 & 0.079 & 0.000 & 0.611 \\ 
        & AquaL. & \cellcolor{gray!15}1.000 & 0.289 & 0.507 & \cellcolor{gray!15}1.000 & \cellcolor{gray!15}1.000 & 0.932 & \cellcolor{gray!15}1.000 & 0.127 & 0.258 & 0.975 & 0.958 & 0.742 & 0.708 \\ 
        & TreeR. & 1.000 & 0.775 & 0.909 & \cellcolor{gray!15}1.000 & \cellcolor{gray!15}1.000 & 0.954 & 0.968 & 0.013 & 0.021 & 0.954 & \cellcolor{gray!15}1.000 & \cellcolor{gray!15}1.000 & 0.778 \\
        & G.S. & \cellcolor{gray!15}1.000 & \cellcolor{gray!15}1.000 & \cellcolor{gray!15}1.000 & \cellcolor{gray!15}1.000 & \cellcolor{gray!15}1.000 & \cellcolor{gray!15}1.000 & \cellcolor{gray!15}1.000 & 0.981 & 0.979 & \cellcolor{gray!15}1.000 & \cellcolor{gray!15}1.000 & \cellcolor{gray!15}1.000 & \cellcolor{gray!15}0.996 \\
        & ROBIN & \cellcolor{gray!15}1.000 & 0.913 & 0.684 & 0.921 & \cellcolor{gray!15}1.000 & 0.013 & 0.935 & \cellcolor{gray!15}1.000 & \cellcolor{gray!15}1.000 & \cellcolor{gray!15}1.000 & \cellcolor{gray!15}1.000 & 0.638 & 0.826 \\ 
        & ZoDiac & \cellcolor{gray!15}1.000 & 0.535 & 0.375 & 0.959 & 0.979 & 0.357 & 0.965 & 0.017 & 0.021 & 0.660 & 0.685 & \cellcolor{gray!15}1.000 & 0.592 \\ 
        & ALIEN-Q & \cellcolor{gray!15}1.000 & 0.783 & \cellcolor{gray!15}1.000 & 0.954 & \cellcolor{gray!15}1.000 & 0.645 & \cellcolor{gray!15}1.000 & 0.153 & 0.311 & 0.965 & 0.970 & 0.945 & 0.793 \\ 
        & ALIEN-R & \cellcolor{gray!15}1.000 & \cellcolor{gray!15}1.000 & \cellcolor{gray!15}1.000 & \cellcolor{gray!15}1.000 & \cellcolor{gray!15}1.000 & 0.979 & \cellcolor{gray!15}1.000 & 0.989 & 0.988 & \cellcolor{gray!15}1.000 & \cellcolor{gray!15}1.000 & \cellcolor{gray!15}1.000 & \cellcolor{gray!15}0.996 \\ 
        \midrule
        
        \multirow{6}{*}{\centering \shortstack{Bit Acc.}} 
        & StegaS. & \cellcolor{gray!15}0.999 & 0.811 & 0.850 & \cellcolor{gray!15}0.999 & \cellcolor{gray!15}0.999 & 0.821 & 0.999 & 0.667 & 0.681 & 0.837 & 0.816 & 0.692 & 0.834 \\ 
        & StableS. & 0.998 & 0.892 & 0.851 & 0.712 & 0.845 & 0.535 & 0.878 & \cellcolor{gray!15}0.991 & \cellcolor{gray!15}0.993 & 0.506 & 0.541 & 0.489 & 0.748 \\ 
        & AquaL. & 0.954 & 0.583 & 0.672 & 0.935 & 0.954 & 0.810 & 0.925 & 0.573 & 0.648 & 0.845 & 0.856 & 0.757 & 0.778 \\ 
        & G.S. & \cellcolor{gray!15}0.999 & 0.939 & 0.964 & 0.995 & 0.\cellcolor{gray!15}999 & 0.903 & 0.996 & 0.657 & 0.655 & \cellcolor{gray!15}0.996 & \cellcolor{gray!15}0.997 & 0.978 & 0.916 \\
        & ALIEN-Q & 0.989 & 0.753 & 0.803 & 0.889 & 0.936 & 0.689 & 0.903 & 0.581 & 0.601 & 0.834 & 0.853 & 0.848 & 0.790 \\ 
        & ALIEN-R & \cellcolor{gray!15}0.999 & \cellcolor{gray!15}0.949 & \cellcolor{gray!15}0.992 & \cellcolor{gray!15}0.999 & \cellcolor{gray!15}0.999 & 0.855 & \cellcolor{gray!15}0.999 & 0.826 & 0.835 & 0.979 & 0.983 & \cellcolor{gray!15}0.999 & \cellcolor{gray!15}0.947 \\ 
        
        \bottomrule
    \end{tabular}
\end{table*}

\textbf{Watermark Stability.} We validate watermark stability across three dimensions: \textit{Sampler Agnosticism} covers both deterministic (DDIM) and irreversible samplers (Euler Ancestral \cite{karras2022elucidating}, DPM++ 2M Ancestral \cite{karras2022elucidating}, DPM++ SDE \cite{lu2025dpm}); \textit{Component Adaptability} involves substituting different VAE and U-Net versions; and \textit{Hyperparameter Stability} examines sensitivity to Guidance Scale and Inference Steps. As shown in Tab.~\ref{tab:stability_bit_acc}, existing training-free methods exhibit vulnerability to stochastic samplers. Specifically, Tree-Ring and Gaussian Shading fail completely under DPM++ SDE, Euler a., and DPM2 a., where accuracy hovers near the random-guess baseline. ALIEN achieves state-of-the-art sampler agnosticism by maintaining high bit accuracy exceeding 0.97 across all tested schedulers. Our method demonstrates consistent stability across varying inference steps, guidance scales, and model versions.
\begin{figure}[t]
  \centering 
\includegraphics[width=1.0\columnwidth]{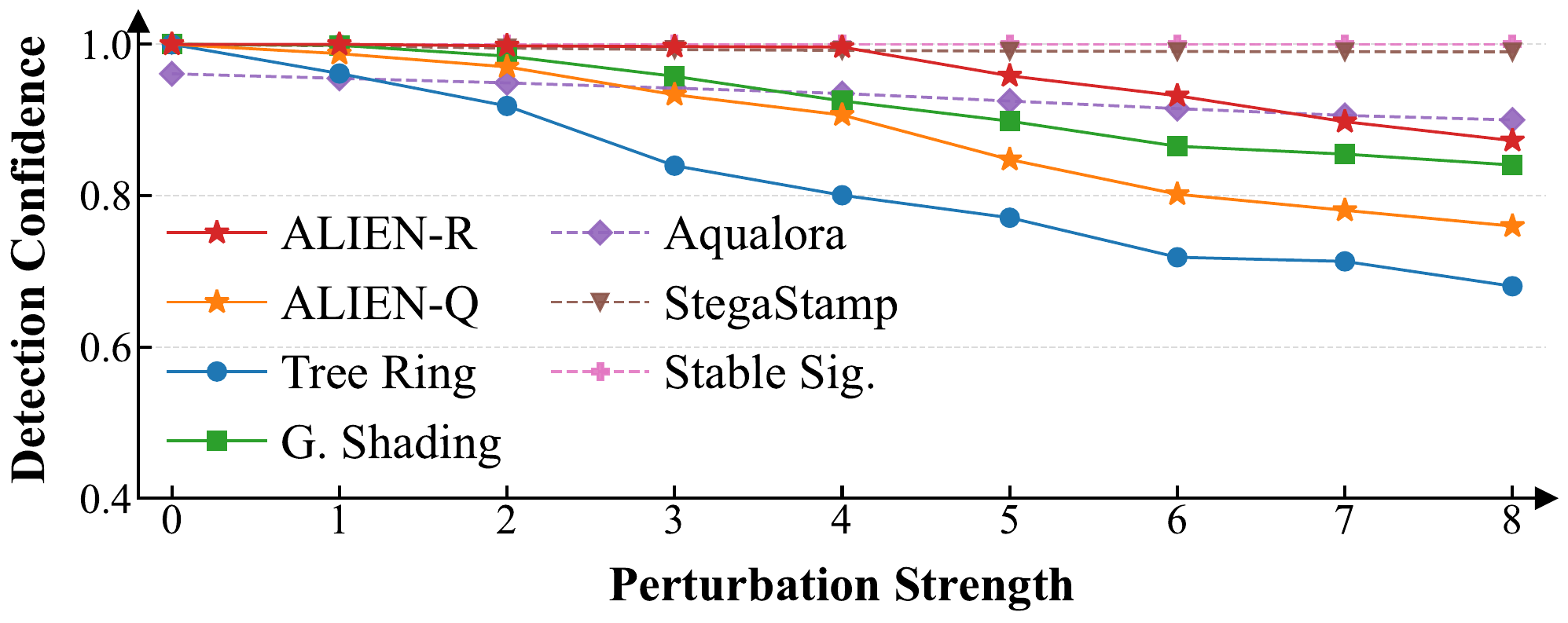}
\caption{Robustness against VAE embedding attacks. Results are averaged across KLVAE8, KLVAE16, and SDXL variants.}
  
  \label{fig:embedding_attack} %
\end{figure}

\textbf{Watermark Efficiency.} We evaluate the computational overhead across two primary dimensions: \textit{Preparation Cost}, the one-time offline cost, and \textit{Runtime Cost}, the online per-image latency for embedding and extraction. The comparative results are detailed in Tab.~\ref{tab:efficiency_comparison_final_no_bold}. ALIEN achieves latencies of 0.079 seconds for embedding and 0.023 seconds for extraction, maintaining speeds comparable to post-processing baselines while being orders of magnitude faster than the optimization-based method Zodiac, which requires over 122 seconds per image.

\textbf{Watermark Robustness.} We examine watermark’s robustness against three typical
kinds of adversarial attacks, including image processing attacks \cite{hore2010image} for transforming generated images, adversarial attacks \cite{an2024waves} for
disturbing watermark verification, and reconstructive attacks \cite{zhao2023invisible} for
re-generating non-watermarked images. Please refer to the Supplementary Materials for detailed parameter settings. As demonstrated in Tab.~\ref{tab:robustness_final_with_avg}, ALIEN-R achieves state-of-the-art stability against standard distortions by maintaining a True Positive Rate of 0.996. In adversarial embedding scenarios illustrated in Fig. \ref{fig:embedding_attack}, ALIEN-R exhibits exceptional resistance where latent semantic watermarking baselines degrade rapidly. Tab.~\ref{tab:robustness_by_scheduler} shows that against reconstructive attacks, ALIEN-R sustains a high confidence exceeding 0.875 even after four re-generation rounds, whereas competitors like StegaStamp and Stable Signature collapse to near-zero.

\textbf{Forgery Resistance.} Following existing research \cite{muller2025black,yang2024can}, we evaluate the security of our method against three representative forgery attacks: imprinting forgery, reprompting forgery, and average forgery. Imprinting and reprompting forgeries rely on latent optimization via diffusion inversion. ALIEN exhibits superior resistance to these attacks compared to initial-latent-based watermarks. While baselines like Tree-Ring and Gaussian Shading can be successfully forged using only a single image, ALIEN remains unaffected. 
Average Forgery involves estimating the watermark pattern by collecting pairs of watermarked and non-watermarked images. While baselines like Tree-Ring exhibit vulnerability with confidence exceeding 0.9, ALIEN maintains a suppressed confidence of 0.711, effectively preventing the rapid extraction of the watermark pattern. The practical challenges in acquiring large-scale specific-user data within real-world API scenarios significantly elevate the security threshold.



\newlength{\nnewidth}
\settowidth{\nnewidth}{00.00}
\newcommand{\nne}{\makebox[\nnewidth]{----}}

\begin{table}[t]
    \centering
    \small
    \setlength{\tabcolsep}{4pt} 
    \caption{Quantitative comparison under Forgery Attacks. Lower values indicate better resistance ($\downarrow$).}
    \label{tab:forgery_combined}
    
    \begin{tabular}{c|c|c|c}
        \toprule
        
        \multirow{2}{*}{\textbf{Config}} & \textbf{G.S.} & \textbf{Tree-Ring} & \textbf{ALIEN (Ours)} \\
        \cmidrule(lr){2-4}
         & \textbf{Acc. / PSNR} & \textbf{Acc. / PSNR} & \textbf{Acc. / PSNR} \\
        \midrule
        
        \multicolumn{4}{c}{\textit{Scenario A: Average Forgery}} \\
        \midrule
        10 Imgs & 0.952 / 12.23 & 0.882 / 18.86 & 0.539 / 20.06 \\
        50 Imgs & 0.969 / 12.79 & 0.908 / 20.34 & 0.605 / 25.29 \\
        100 Imgs & 0.969 / 12.65 & 0.910 / 20.95 & 0.708 / 26.48 \\
        \midrule
        
        \multicolumn{4}{c}{\textit{Scenario B: Reprompt Forgery}} \\
        \midrule
        SD-V2.1 & 1.000 / \nne & 0.931 / \nne & 0.533 / \nne \\
        \bottomrule
    \end{tabular}
    \vspace{-0.2em}
\end{table}
\begin{figure}[t]
  \begin{center}
\centerline{\includegraphics[width=1\columnwidth]{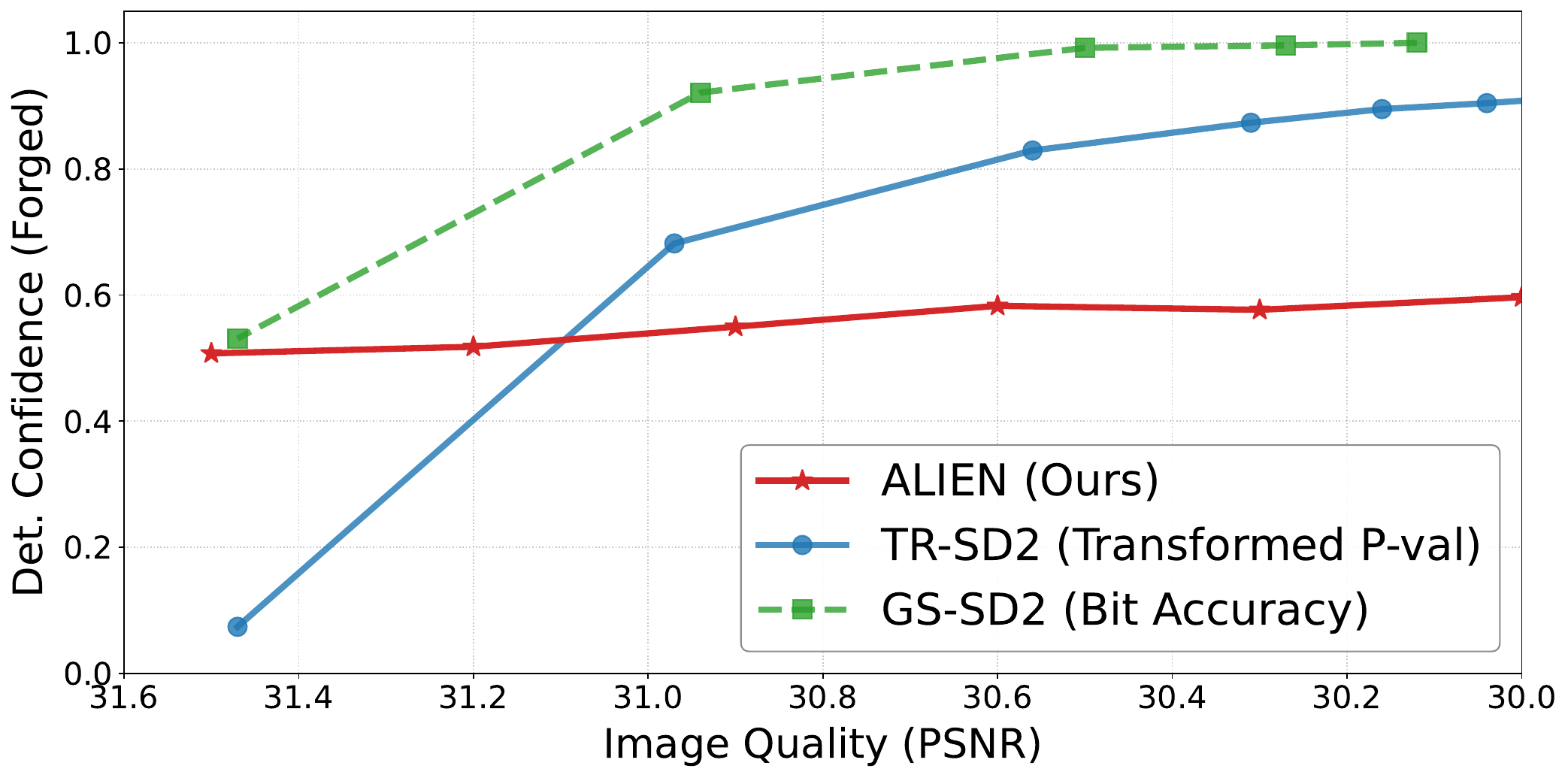}}
    \caption{Robustness against Imprinting Forgery Attack.}
    \label{forge}
  \end{center}
  \vspace{-0.4em}
\end{figure}

        
        

\begin{table}[t]
    \centering
    \small
    \setlength{\tabcolsep}{3pt} 
    \caption{Ablation study on Timestep Range under different Watermark Strengths. We compare Image Quality (PSNR) and Detection Performance (Det./Acc.) across two strength settings.}
    \label{tab:ablation_range_strength}
    
    \begin{tabular}{c|cc|cc}
        \toprule
        \multirow{2}{*}{\textbf{Range}} & \multicolumn{2}{c|}{\textbf{Strength A} ($\lambda=0.5$)} & \multicolumn{2}{c}{\textbf{Strength B} ($\lambda=1.0$)} \\
        \cmidrule(lr){2-3} \cmidrule(lr){4-5}
        ($t_{start} \to t_{end}$) & \textbf{PSNR} & \textbf{Det. / Acc.} & \textbf{PSNR} & \textbf{Det. / Acc.} \\
        \midrule
        
        $1 \to 50$  & 24.55 & 1.000 / 1.000 & 20.42 & 1.000 / 1.000 \\
        $25 \to 50$ & 30.53 & 1.000 / 1.000 & 26.27 & 1.000 / 1.000 \\
        $25 \to 40$  & 40.94 & 0.718 / 0.747 & 23.10 & 0.852 / 1.000 \\
        $10 \to 25$ & 31.54 & 0.614 / 0.421 & 26.83 & 0.681 / 0.721 \\
        $20 \to 45$ & 33.76 & 0.882 / 1.000 & 31.10 & 0.989 / 1.000 \\
        \bottomrule
    \end{tabular}
\end{table}

\textbf{Impact of Timestep Range.} As shown in Tab.~\ref{tab:ablation_range_strength}, we conducted an ablation study on the injection interval to investigate the trade-off between imperceptibility and robustness. Injecting the watermark across the full reverse process from step 50 down to 1 ensures perfect detection with 1.000 accuracy but introduces significant perceptual degradation, resulting in a PSNR drop to 20.42 dB under Strength B. Narrowing the window to early diffusion stages such as steps 45 to 20 significantly recovers image quality, where the PSNR increases to 33.76 dB under Strength A while maintaining a high detection rate.

\textbf{Impact of Watermark Strength.} As shown in Tab.~\ref{tab:ablation_range_strength}, the injection coefficient $\lambda$ directly modulates the magnitude of the probability flow correction. Increasing the strength from $\lambda=0.5$ in Strength A to $\lambda=1.0$ in Strength B consistently improves detection performance across all ranges. we select the range from 45 to 20 combined with moderate strength as the optimal configuration for ALIEN-Q to maximize quality while ensuring reliable detection.

\begin{table}[t]
    \centering
    \caption{Performance Comparison regarding Computational Efficiency. $^{\ddagger}$ indicates data estimated from original papers.}
    \scriptsize \label{tab:efficiency_comparison_final_no_bold}
    \setlength{\tabcolsep}{3pt} 
    \footnotesize
    \begin{tabular}{ccccccc}
        \toprule
        \multirow{2}{*}{\centering Method} & \multicolumn{3}{c}{Prep. Time (h)} & Embed. & Extract. \\
        \cmidrule(lr){2-4}
        & Pre-train & Fine-tune & Optimize & Time (s) & Time (s) \\
        \midrule
        
        T.R. & --- & --- & --- & 0.011 & 3.575 \\
        G.S. & --- & --- & --- & 0.102& 4.212\\
        ROBIN & --- & --- & 0.4 & 0.125& 1.248 \\
        Zodiac & --- & --- & --- & 122.4 & 3.601 \\
        StegaS. & 30 & --- & --- & 0.108 & 0.028 \\
        S.S. & 48$^{\ddagger}$ & 0.1 & --- & --- & 0.014 \\
        AquaL. & 40$^{\ddagger}$ & 15$^{\ddagger}$ & --- & --- & 0.026 \\
        \midrule
        ALIEN & 22 & --- & --- & 0.079 & 0.023 \\
        \bottomrule
    \end{tabular}
\end{table}

\section{Conclusion}

Unlike prior approaches relying on heuristic latent modifications, sampling constraints, or computationally intensive optimization, we propose ALIEN, which presents the first analytical derivation of the time-dependent modulation coefficient. This principled approach precisely guides the diffusion of watermark residuals via probability flow modulation, achieving sampler-agnostic embedding. ALIEN overcomes the security vulnerabilities associated with diffusion inversion and irreversible samplers while ensuring high fidelity and low inference cost. Experimental results demonstrate that ALIEN outperforms existing methods in both quality and robustness. Future research will focus on developing content-aware latent watermarking to further enhance security against forgery attacks.

\section{Impact Statement}
Ensuring the accountability and intellectual property protection of generative models has become a critical priority. This paper introduces ALIEN, a principled watermarking framework that enables robust and sampler-agnostic provenance tracking for Latent Diffusion Models. By overcoming the vulnerability of existing methods to irreversible samplers and diverse attacks, we believe our approach effectively mitigates risks such as copyright infringement and malicious misuse, thereby contributing to the development of safer and more trustworthy generative AI.

\bibliography{example_paper}

@article{ramesh2022hierarchical,
  title={Hierarchical text-conditional image generation with clip latents},
  author={Ramesh, Aditya and Dhariwal, Prafulla and Nichol, Alex and Chu, Casey and Chen, Mark},
  journal={arXiv preprint arXiv:2204.06125},
  volume={1},
  number={2},
  pages={3},
  year={2022}
}

@article{barrett2023identifying,
  title={Identifying and mitigating the security risks of generative ai},
  author={Barrett, Clark},
  journal={arXiv preprint arXiv:2308.14840},
  year={2023}
}

@misc{gowal2023identifying,
  author = {Sven Gowal and Pushmeet Kohli},
  title = {Identifying AI-generated images with SynthID},
  year = {2023},
  howpublished = {\url{https://www.deepmind.com/blog/identifying-ai-generated-images-with-synthid}},
  note = {Accessed: 2023-09-23}
}

@inproceedings{fernandez2023stable,
  title={The stable signature: Rooting watermarks in latent diffusion models},
  author={Fernandez, Pierre and Couairon, Guillaume and J{\'e}gou, Herv{\'e} and Douze, Matthijs and Furon, Teddy},
  booktitle={Proceedings of the IEEE/CVF International Conference on Computer Vision},
  pages={22466--22477},
  year={2023}
}

@article{zhao2023invisible,
  title={Invisible image watermarks are provably removable using generative ai},
  author={Zhao, Xuandong and Zhang, Kexun and Su, Zihao and Vasan, Saastha and Grishchenko, Ilya and Kruegel, Christopher and Vigna, Giovanni and Wang, Yu-Xiang and Li, Lei},
  journal={arXiv preprint arXiv:2306.01953},
  year={2023}
}

@article{wen2024tree,
  title={Tree-rings watermarks: Invisible fingerprints for diffusion images},
  author={Wen, Yuxin and Kirchenbauer, John and Geiping, Jonas and Goldstein, Tom},
  journal={Advances in Neural Information Processing Systems},
  volume={36},
  year={2024}
}

@article{ho2020denoising,
  title={Denoising diffusion probabilistic models},
  author={Ho, Jonathan and Jain, Ajay and Abbeel, Pieter},
  journal={Advances in neural information processing systems},
  volume={33},
  pages={6840--6851},
  year={2020}
}

@article{song2020denoising,
  title={Denoising diffusion implicit models},
  author={Song, Jiaming and Meng, Chenlin and Ermon, Stefano},
  journal={arXiv preprint arXiv:2010.02502},
  year={2020}
}

@article{song2020score,
  title={Score-based generative modeling through stochastic differential equations},
  author={Song, Yang and Sohl-Dickstein, Jascha and Kingma, Diederik P and Kumar, Abhishek and Ermon, Stefano and Poole, Ben},
  journal={arXiv preprint arXiv:2011.13456},
  year={2020}
}

@inproceedings{tancik2020stegastamp,
  title={Stegastamp: Invisible hyperlinks in physical photographs},
  author={Tancik, Matthew and Mildenhall, Ben and Ng, Ren},
  booktitle={Proceedings of the IEEE/CVF Conference on Computer Vision and Pattern Recognition},
  pages={2117--2126},
  year={2020}
}

@article{meng2024latent,
  title={Latent Watermark: Inject and Detect Watermarks in Latent Diffusion Space},
  author={Meng, Zheling and Peng, Bo and Dong, Jing},
  journal={arXiv preprint arXiv:2404.00230},
  year={2024}
}

@article{dhariwal2021diffusion,
  title={Diffusion models beat gans on image synthesis},
  author={Dhariwal, Prafulla and Nichol, Alexander},
  journal={Advances in neural information processing systems},
  volume={34},
  pages={8780--8794},
  year={2021}
}

@inproceedings{lin2014microsoft,
  title={Microsoft coco: Common objects in context},
  author={Lin, Tsung-Yi and Maire, Michael and Belongie, Serge and Hays, James and Perona, Pietro and Ramanan, Deva and Doll{\'a}r, Piotr and Zitnick, C Lawrence},
  booktitle={Computer Vision--ECCV 2014: 13th European Conference, Zurich, Switzerland, September 6-12, 2014, Proceedings, Part V 13},
  pages={740--755},
  year={2014},
  organization={Springer}
}

@article{balle2018variational,
  title={Variational image compression with a scale hyperprior},
  author={Ball{\'e}, Johannes and Minnen, David and Singh, Saurabh and Hwang, Sung Jin and Johnston, Nick},
  journal={arXiv preprint arXiv:1802.01436},
  year={2018}
}

@inproceedings{cheng2020learned,
  title={Learned image compression with discretized gaussian mixture likelihoods and attention modules},
  author={Cheng, Zhengxue and Sun, Heming and Takeuchi, Masaru and Katto, Jiro},
  booktitle={Proceedings of the IEEE/CVF Conference on Computer Vision and Pattern Recognition},
  pages={7939--7948},
  year={2020}
}

@article{song2020improved,
  title={Improved techniques for training score-based generative models},
  author={Song, Yang and Ermon, Stefano},
  journal={Advances in neural information processing systems},
  volume={33},
  pages={12438--12448},
  year={2020}
}

@book{cox2007digital,
  title={Digital watermarking and steganography},
  author={Cox, Ingemar and Miller, Matthew and Bloom, Jeffrey and Fridrich, Jessica and Kalker, Ton},
  year={2007},
  publisher={Morgan kaufmann}
}

@inproceedings{hore2010image,
  title={Image quality metrics: PSNR vs. SSIM},
  author={Hore, Alain and Ziou, Djemel},
  booktitle={2010 20th international conference on pattern recognition},
  pages={2366--2369},
  year={2010},
  organization={IEEE}
}

@article{wang2004image,
  title={Image quality assessment: from error visibility to structural similarity},
  author={Wang, Zhou and Bovik, Alan C and Sheikh, Hamid R and Simoncelli, Eero P},
  journal={IEEE transactions on image processing},
  volume={13},
  number={4},
  pages={600--612},
  year={2004},
  publisher={IEEE}
}

@article{heusel2017gans,
  title={Gans trained by a two time-scale update rule converge to a local nash equilibrium},
  author={Heusel, Martin and Ramsauer, Hubert and Unterthiner, Thomas and Nessler, Bernhard and Hochreiter, Sepp},
  journal={Advances in neural information processing systems},
  volume={30},
  year={2017}
}

@inproceedings{Radford2021LearningTV,
  title={Learning Transferable Visual Models From Natural Language Supervision},
  author={Alec Radford and Jong Wook Kim and Chris Hallacy and A. Ramesh and Gabriel Goh and Sandhini Agarwal and Girish Sastry and Amanda Askell and Pamela Mishkin and Jack Clark and Gretchen Krueger and Ilya Sutskever},
  booktitle={ICML},
  year={2021}
}

@article{lu2022dpm,
  title={Dpm-solver: A fast ode solver for diffusion probabilistic model sampling in around 10 steps},
  author={Lu, Cheng and Zhou, Yuhao and Bao, Fan and Chen, Jianfei and Li, Chongxuan and Zhu, Jun},
  journal={Advances in Neural Information Processing Systems},
  volume={35},
  pages={5775--5787},
  year={2022}
}

@article{karras2022elucidating,
  title={Elucidating the design space of diffusion-based generative models},
  author={Karras, Tero and Aittala, Miika and Aila, Timo and Laine, Samuli},
  journal={Advances in neural information processing systems},
  volume={35},
  pages={26565--26577},
  year={2022}
}

@InProceedings{Rombach_2022_CVPR,
    author    = {Rombach, Robin and Blattmann, Andreas and Lorenz, Dominik and Esser, Patrick and Ommer, Bj\"orn},
    title     = {High-Resolution Image Synthesis With Latent Diffusion Models},
    booktitle = {Proceedings of the IEEE/CVF Conference on Computer Vision and Pattern Recognition (CVPR)},
    month     = {June},
    year      = {2022},
    pages     = {10684-10695}
}

@article{feng2024aqualora,
  title={AquaLoRA: Toward White-box Protection for Customized Stable Diffusion Models via Watermark LoRA},
  author={Feng, Weitao and Zhou, Wenbo and He, Jiyan and Zhang, Jie and Wei, Tianyi and Li, Guanlin and Zhang, Tianwei and Zhang, Weiming and Yu, Nenghai},
  journal={arXiv preprint arXiv:2405.11135},
  year={2024}
}

@inproceedings{yang2024gaussian,
  title={Gaussian Shading: Provable Performance-Lossless Image Watermarking for Diffusion Models},
  author={Yang, Zijin and Zeng, Kai and Chen, Kejiang and Fang, Han and Zhang, Weiming and Yu, Nenghai},
  booktitle={Proceedings of the IEEE/CVF Conference on Computer Vision and Pattern Recognition},
  pages={12162--12171},
  year={2024}
}

@inproceedings{ci2025ringid,
  title={Ringid: Rethinking tree-ring watermarking for enhanced multi-key identification},
  author={Ci, Hai and Yang, Pei and Song, Yiren and Shou, Mike Zheng},
  booktitle={European Conference on Computer Vision},
  pages={338--354},
  year={2025},
  organization={Springer}
}

@misc{EU_AI_Act_2024,
  author       = {{European Union}},
  title        = {Artificial Intelligence Act: Regulation (EU) 2024/1689 of the European Parliament and of the Council},
  year         = {2024},
  month        = jun,
  note         = {URL: \url{https://eur-lex.europa.eu/legal-content/EN/TXT/?uri=CELEX:32024R1689}, Accessed: 2024-09-24},
}

@misc{Biden_2023,
  author       = {Biden, J. R.},
  title        = {Executive order on the safe, secure, and trustworthy development and use of artificial intelligence},
  year         = {2023},
  url = {https://www.whitehouse.gov/briefing-room/statements-releases/2023/07/07/fact-sheet-executive-order-on-artificial-intelligence/},
}

@article{ren2024sok,
  title={Sok: On the role and future of aigc watermarking in the era of gen-ai},
  author={Ren, Kui and Yang, Ziqi and Lu, Li and Liu, Jian and Li, Yiming and Wan, Jie and Zhao, Xiaodi and Feng, Xianheng and Shao, Shuo},
  journal={arXiv preprint arXiv:2411.11478},
  year={2024}
}

@inproceedings{rezaei2024lawa,
  title={Lawa: Using latent space for in-generation image watermarking},
  author={Rezaei, Ahmad and Akbari, Mohammad and Alvar, Saeed Ranjbar and Fatemi, Arezou and Zhang, Yong},
  booktitle={European Conference on Computer Vision},
  pages={118--136},
  year={2024},
  organization={Springer}
}

@article{ci2024wmadapter,
  title={Wmadapter: Adding watermark control to latent diffusion models},
  author={Ci, Hai and Song, Yiren and Yang, Pei and Xie, Jinheng and Shou, Mike Zheng},
  journal={arXiv preprint arXiv:2406.08337},
  year={2024}
}

@inproceedings{wang2025sleepermark,
  title={Sleepermark: Towards robust watermark against fine-tuning text-to-image diffusion models},
  author={Wang, Zilan and Guo, Junfeng and Zhu, Jiacheng and Li, Yiming and Huang, Heng and Chen, Muhao and Tu, Zhengzhong},
  booktitle={Proceedings of the Computer Vision and Pattern Recognition Conference},
  pages={8213--8224},
  year={2025}
}

@inproceedings{lee2025semantic,
  title={Semantic Watermarking Reinvented: Enhancing Robustness and Generation Quality with Fourier Integrity},
  author={Lee, Sung Ju and Cho, Nam Ik},
  booktitle={Proceedings of the IEEE/CVF International Conference on Computer Vision},
  pages={18759--18769},
  year={2025}
}

@article{zhao2024invisible,
  title={Invisible image watermarks are provably removable using generative ai},
  author={Zhao, Xuandong and Zhang, Kexun and Su, Zihao and Vasan, Saastha and Grishchenko, Ilya and Kruegel, Christopher and Vigna, Giovanni and Wang, Yu-Xiang and Li, Lei},
  journal={Advances in neural information processing systems},
  volume={37},
  pages={8643--8672},
  year={2024}
}

@article{gunn2024undetectable,
  title={An undetectable watermark for generative image models},
  author={Gunn, Sam and Zhao, Xuandong and Song, Dawn},
  journal={arXiv preprint arXiv:2410.07369},
  year={2024}
}

@article{huang2024robin,
  title={Robin: Robust and invisible watermarks for diffusion models with adversarial optimization},
  author={Huang, Huayang and Wu, Yu and Wang, Qian},
  journal={Advances in Neural Information Processing Systems},
  volume={37},
  pages={3937--3963},
  year={2024}
}

@article{zhang2024attack,
  title={Attack-resilient image watermarking using stable diffusion},
  author={Zhang, Lijun and Liu, Xiao and Martin, Antoni V and Bearfield, Cindy X and Brun, Yuriy and Guan, Hui},
  journal={Advances in Neural Information Processing Systems},
  volume={37},
  pages={38480--38507},
  year={2024}
}

@inproceedings{muller2025black,
  title={Black-box forgery attacks on semantic watermarks for diffusion models},
  author={M{\"u}ller, Andreas and Lukovnikov, Denis and Thietke, Jonas and Fischer, Asja and Quiring, Erwin},
  booktitle={Proceedings of the Computer Vision and Pattern Recognition Conference},
  pages={20937--20946},
  year={2025}
}

@article{li2025gaussmarker,
  title={GaussMarker: Robust Dual-Domain Watermark for Diffusion Models},
  author={Li, Kecen and Huang, Zhicong and Hou, Xinwen and Hong, Cheng},
  journal={arXiv preprint arXiv:2506.11444},
  year={2025}
}

@inproceedings{zhang2018unreasonable,
  title={The unreasonable effectiveness of deep features as a perceptual metric},
  author={Zhang, Richard and Isola, Phillip and Efros, Alexei A and Shechtman, Eli and Wang, Oliver},
  booktitle={Proceedings of the IEEE conference on computer vision and pattern recognition},
  pages={586--595},
  year={2018}
}

@dataset{Gustavosta2023,
  author = {Gustavosta},
  title = {Stable Diffusion Prompts Dataset},
  year = {2023},
  month = may,
  url = {https://huggingface.co/datasets/Gustavosta/Stable-Diffusion-Prompts},
  version = {1.0},
  publisher = {Hugging Face}
}

@article{an2024waves,
  title={Waves: Benchmarking the robustness of image watermarks},
  author={An, Bang and Ding, Mucong and Rabbani, Tahseen and Agrawal, Aakriti and Xu, Yuancheng and Deng, Chenghao and Zhu, Sicheng and Mohamed, Abdirisak and Wen, Yuxin and Goldstein, Tom and others},
  journal={arXiv preprint arXiv:2401.08573},
  year={2024}
}

@article{fu2023dreamsim,
  title={Dreamsim: Learning new dimensions of human visual similarity using synthetic data},
  author={Fu, Stephanie and Tamir, Netanel and Sundaram, Shobhita and Chai, Lucy and Zhang, Richard and Dekel, Tali and Isola, Phillip},
  journal={arXiv preprint arXiv:2306.09344},
  year={2023}
}

@article{yang2024sifid,
  title={Sifid: Reassess summary factual inconsistency detection with llm},
  author={Yang, Jiuding and Liu, Hui and Guo, Weidong and Rao, Zhuwei and Xu, Yu and Niu, Di},
  journal={arXiv preprint arXiv:2403.07557},
  year={2024}
}

@article{lu2025dpm,
  title={Dpm-solver++: Fast solver for guided sampling of diffusion probabilistic models},
  author={Lu, Cheng and Zhou, Yuhao and Bao, Fan and Chen, Jianfei and Li, Chongxuan and Zhu, Jun},
  journal={Machine Intelligence Research},
  pages={1--22},
  year={2025},
  publisher={Springer}
}

@article{yang2024can,
  title={Can simple averaging defeat modern watermarks?},
  author={Yang, Pei and Ci, Hai and Song, Yiren and Shou, Mike Zheng},
  journal={Advances in Neural Information Processing Systems},
  volume={37},
  pages={56644--56673},
  year={2024}
}
\bibliographystyle{icml2026}

\newpage
\appendix
\appendix
\onecolumn

\clearpage
\thispagestyle{plain}

\newcommand{\maintitlefont}{\large}
\newcommand{\subtitlefont}{\large}
\newcommand{\pagespacing}{0.5em}
\newcommand{\sectionspacing}{1em}
\newcommand{\dotfilll}{\leaders\hbox to 0.5em{\hfil.\hfil}\hfill}

\begin{center}
    \LARGE\textbf{Supplementary Material}
\end{center}

\vspace{1cm}

\noindent
\begin{minipage}{\textwidth}
\setlength{\parindent}{0pt}

\noindent
\maintitlefont\hyperref[app:theory]{\textbf{A \hspace{0.5em} Theoretical Analysis of ALIEN} \hfill \textbf{\pageref{app:theory}}} \\[\pagespacing]
\subtitlefont\hyperref[app:wm_constraint]{\hspace{1.5em} A.1 \hspace{0.5em} Watermark Constraint and Score Discrepancy \dotfilll \pageref{app:wm_constraint}} \\[\pagespacing]
\subtitlefont\hyperref[app:sde_drift]{\hspace{1.5em} A.2 \hspace{0.5em} Proof of Reverse SDE Drift Term Correction \dotfilll \pageref{app:sde_drift}} \\[\pagespacing]
\subtitlefont\hyperref[app:unet_target]{\hspace{1.5em} A.3 \hspace{0.5em} Derivation of UNet Target Noise \dotfilll \pageref{app:unet_target}}

\par\vspace{\sectionspacing}

\noindent
\maintitlefont\hyperref[app:impl]{\textbf{B \hspace{0.5em} Implementation Details} \hfill \textbf{\pageref{app:impl}}} \\[\pagespacing]
\subtitlefont\hyperref[app:pseudocode]{\hspace{1.5em} B.1 \hspace{0.5em} Algorithm Pseudocode \dotfilll \pageref{app:pseudocode}} \\[\pagespacing]
\subtitlefont\hyperref[app:training_details]{\hspace{1.5em} B.2 \hspace{0.5em} Hyperparameters and Training Settings \dotfilll \pageref{app:training_details}}

\par\vspace{\sectionspacing}

\noindent
\maintitlefont\hyperref[app:setup]{\textbf{C \hspace{0.5em} Experimental Setup and Extended Experiments} \hfill \textbf{\pageref{app:setup}}} \\[\pagespacing]
\subtitlefont\hyperref[app:datasets]{\hspace{1.5em} C.1 \hspace{0.5em} Dataset Settings \dotfilll \pageref{app:datasets}} \\[\pagespacing]
\subtitlefont\hyperref[app:robustness_settings]{\hspace{1.5em} C.2 \hspace{0.5em} Robustness Evaluation Settings \dotfilll \pageref{app:robustness_settings}} \\[\pagespacing]
\subtitlefont\hyperref[app:add_experiments]{\hspace{1.5em} C.3 \hspace{0.5em} Extended Experiments \dotfilll \pageref{app:add_experiments}}

\par\vspace{\sectionspacing}

\noindent
\maintitlefont\hyperref[app:related]{\textbf{D \hspace{0.5em} Extended Related Work} \hfill \textbf{\pageref{app:related}}} \\[\pagespacing]
\subtitlefont\hyperref[app:taxonomy]{\hspace{1.5em} D.1 \hspace{0.5em} Watermarking Schemes Taxonomy \dotfilll \pageref{app:taxonomy}} \\[\pagespacing]

\end{minipage}
\clearpage
\normalsize


\section{Theoretical Analysis of ALIEN}
\label{app:theory}
In this section, we provide a rigorous derivation establishing the analytical link between the watermark constraint in the $\mathbf{z}_0$-space and the necessary correction to the probability flow drift term. This derivation proves that ALIEN is theoretically grounded in the VP-SDE framework and is inherently sampler-agnostic.

\subsection{Watermark Constraint and Score Function Discrepancy}
\label{app:wm_constraint}
\paragraph{1. Watermark Constraint Definition}
Our objective is to embed a fixed watermark residual $\mathbf{\delta}_{wm}$ into the latent representation. We define this as a geometric constraint on the estimated clean data manifold. Let $\hat{\mathbf{z}}_0$ denote the original estimate derived from the model $\theta$, and $\hat{\mathbf{z}}_0^{wm}$ denote the target watermarked estimate:
\begin{equation}
    \hat{\mathbf{z}}_0^{wm} = \hat{\mathbf{z}}_0 + \mathbf{\delta}_{wm} \label{eq:wm_constraint}.
\end{equation}

\paragraph{2. Derivation of Score Function Difference $\Delta \text{Score}$}
Under the VP-SDE framework, Tweedie's formula establishes a linear bijection between the score function $\nabla_{\mathbf{z}_t} \log p_t(\mathbf{z}_t)$ and the denoised estimate $\hat{\mathbf{z}}_0$:
\begin{equation}
    \nabla_{\mathbf{z}_t} \log p_t(\mathbf{z}_t) = - \frac{\mathbf{z}_t - \sqrt{\bar{\alpha}_t} \hat{\mathbf{z}}_0}{1 - \bar{\alpha}_t}.
\end{equation}
To enforce the watermark constraint (Eq. \ref{eq:wm_constraint}), the score function must shift to $\nabla_{\mathbf{z}_t} \log p_t^{wm}$. By substituting the constraint $\hat{\mathbf{z}}_0 - \hat{\mathbf{z}}_0^{wm} = -\mathbf{\delta}_{wm}$, we derive the score discrepancy $\Delta \text{Score}$:
\begin{equation}
    \begin{aligned}
    \Delta \text{Score} &= \nabla_{\mathbf{z}_t} \log p_t^{wm} - \nabla_{\mathbf{z}_t} \log p_t \\
    &= -\frac{1}{1 - \bar{\alpha}_t} \left( -\sqrt{\bar{\alpha}_t} (\hat{\mathbf{z}}_0^{wm} - \hat{\mathbf{z}}_0) \right) \\
    &= \frac{\sqrt{\bar{\alpha}_t}}{1 - \bar{\alpha}_t} \mathbf{\delta}_{wm}.
    \end{aligned}
    \label{eq:delta_score}
\end{equation}

\subsection{Proof of Reverse SDE Drift Term Correction $\Delta \mathbf{F}_{rev}$}
\label{app:sde_drift}
\paragraph{Definition of Reverse Drift $\mathbf{F}_{rev}$}
The generation trajectory in diffusion models is governed by the reverse-time SDE. The deterministic component of this process, known as the drift term $\mathbf{F}_{rev}$, is defined as:
\begin{equation}
    \mathbf{F}_{rev}(\mathbf{z}_t, t) = \mathbf{f}(\mathbf{z}_t, t) - g^2(t) \nabla_{\mathbf{z}_t} \log p_t(\mathbf{z}_t).
\end{equation}
where $\mathbf{f}(\mathbf{z}_t, t)$ is the forward drift and $g(t)$ is the diffusion coefficient.

\paragraph{Derivation of Drift Correction $\Delta \mathbf{F}_{rev}$}
We calculate the modification to the drift term required to accommodate the watermark. Since the forward physics $\mathbf{f}(\mathbf{z}_t, t)$ remains invariant, the drift correction depends solely on the score shift:
\begin{equation}
    \Delta \mathbf{F}_{rev} = \mathbf{F}_{rev}^{wm} - \mathbf{F}_{rev}^{orig} = - g^2(t) \left( \Delta \text{Score} \right).
\end{equation}
Substituting Eq. (\ref{eq:delta_score}) into the expression above, we obtain the explicit form of the Watermark Drift Force:
\begin{equation}
    \Delta \mathbf{F}_{rev}(\mathbf{z}_t, t) = - g^2(t) \frac{\sqrt{\bar{\alpha}_t}}{1 - \bar{\alpha}_t} \mathbf{\delta}_{wm}.
\end{equation}
\textbf{Theorem:} The spatial constraint $\mathbf{\delta}_{wm}$ imposes a constant, deterministic force $\Delta \mathbf{F}_{rev}$ on the probability flow. Because $\mathbf{F}_{rev}$ is the shared driving term for both the stochastic reverse SDE and the deterministic Probability Flow ODE (PF-ODE), this correction guarantees that the watermark embedding is robust across different samplers.

\subsection{Derivation of Noise Prediction Target $\mathbf{\epsilon}^{target}$}
\label{app:unet_target}
To implement the theoretical drift correction $\Delta \mathbf{F}_{rev}$ in practice, we modulate the output of the U-Net $\vartheta$.

\paragraph{Score-Noise Relationship}
The neural network $\mathbf{\epsilon}_{\vartheta}$ approximates the score function via the relation:
\begin{equation}
    \nabla_{\mathbf{z}_t} \log p_t(\mathbf{z}_t) = -\frac{\mathbf{\epsilon}_{\vartheta}(\mathbf{z}_t, t)}{\sqrt{1 - \bar{\alpha}_t}}.
\end{equation}

\paragraph{Mapping Drift Correction to $\epsilon$-Space}
We equate the score difference derived from the drift requirement to the difference in noise prediction. Using $\Delta \text{Score} = -\frac{1}{\sqrt{1 - \bar{\alpha}_t}} (\mathbf{\epsilon}^{target} - \mathbf{\epsilon}_{\vartheta}) = -\frac{\Delta \mathbf{\epsilon}}{\sqrt{1 - \bar{\alpha}_t}}$, we have:
\begin{equation}
    -\frac{\Delta \mathbf{\epsilon}}{\sqrt{1 - \bar{\alpha}_t}} = \frac{\sqrt{\bar{\alpha}_t}}{1 - \bar{\alpha}_t} \mathbf{\delta}_{wm}.
\end{equation}

\paragraph{Final Update Rule}
Solving for the correction term $\Delta \mathbf{\epsilon}$:
\begin{equation}
    \Delta \mathbf{\epsilon} = -\sqrt{1 - \bar{\alpha}_t} \cdot \frac{\sqrt{\bar{\alpha}_t}}{1 - \bar{\alpha}_t} \mathbf{\delta}_{wm} = -\frac{\sqrt{\bar{\alpha}_t}}{\sqrt{1 - \bar{\alpha}_t}} \mathbf{\delta}_{wm}.
\end{equation}
Thus, the final target noise $\mathbf{\epsilon}^{target}$ required to enforce the watermark is:
\begin{equation}
     \mathbf{\epsilon}^{target} = \mathbf{\epsilon}_{\vartheta} + \Delta \mathbf{\epsilon} = \mathbf{\epsilon}_{\vartheta} - \left( \frac{\sqrt{\bar{\alpha}_t}}{\sqrt{1 - \bar{\alpha}_t}} \right) \mathbf{\delta}_{wm}.
\end{equation}

\section{Implementation Details}
\label{app:impl}


\begin{algorithm}[t]
\caption{ALIEN Watermarking}
\label{alg:alien}
\begin{algorithmic}[1]

\STATE \textcolor{gray}{\hrulefill\ \textbf{Phase I: Watermark Embedding}\ \hrulefill}

\STATE \textbf{Input}: Pre-trained U-Net $\Unet$, Encoder $\Es$, Scheduler $S$, VAE Decoder $\DVAE$, Prompt $\mathbf{c}$, Secret $\secret$, Injection Interval $[T_{start}, T_{end}]$, Strength $\lambda$
\STATE \textbf{Output}: Watermarked Image $\ximg_{wm}$

\STATE $\wmres \leftarrow \Es(\secret)$
\STATE $\zstep \sim \mathcal{N}(\mathbf{0}, \mathbf{I})$

\STATE \textbf{for} $t = T$ \textbf{down to} $1$ \textbf{do}
    \STATE \quad $\mathbf{t} \leftarrow S.\text{timesteps}[t]$
    
    \STATE \quad $\epred \leftarrow \text{CFG}(\Unet, \zstep, \mathbf{t}, \mathbf{c})$
    
    \STATE \quad \textbf{if} $t \le T_{end}$ \textbf{and} $t \ge T_{start}$ \textbf{then}
        
        \vspace{0.2em} 
        \STATE \quad \quad \textcolor{teal!80!black}{\footnotesize \textit{// Modulate the prediction target}}
        \STATE \quad \quad $\epred \leftarrow \epred - \lambda \cdot \left(\frac{\sqrt{\bar{\alpha}_t}}{\sqrt{1 - \bar{\alpha}_t}}\right) \cdot \wmres$
        \vspace{0.1em}

    \STATE \quad \textbf{end if}

    \STATE \quad $\zstep \leftarrow S.\text{step}(\epred, \mathbf{t}, \zstep).\text{prev\_sample}$
\STATE \textbf{end for}
\STATE $\ximg_{wm} \leftarrow \DVAE(\zstep)$

\STATE \textcolor{gray}{\hrulefill\ \textbf{Phase II: Watermark Extraction}\ \hrulefill}

\STATE \textbf{Input}: Watermarked Image $\ximg_{wm}$, VAE Encoder $\EVAE$, Watermark Decoder $\Ds$
\STATE \textbf{Output}: Extracted Secret $\secret^{\prime}$

\STATE \textcolor{teal!80!black}{\footnotesize \textit{// Encode image back to watermarked latent}}
\STATE $\zfinal \leftarrow \EVAE(\ximg_{wm})$
\STATE \textcolor{teal!80!black}{\footnotesize \textit{// Extract secret message from latent}}
\STATE $\secret^{\prime} \leftarrow \Ds(\zfinal)$
\STATE \textbf{Return} $\secret^{\prime}$

\end{algorithmic}

\end{algorithm}
\subsection{Algorithm Pseudocode}
\label{app:pseudocode}
The core logic of the ALIEN watermarking framework is detailed in Algorithm \ref{alg:alien}. This algorithm outlines the complete process of watermark embedding during the reverse diffusion process and the subsequent extraction from the latent space.
\begin{table}[t]
\centering
\small 
\renewcommand{\arraystretch}{1.1} 
\setlength{\tabcolsep}{4pt} 
\caption{\textbf{Detailed Robustness Evaluation Settings.} We categorize attacks into four distinct groups. Note that Image Processing attacks are grouped by type (Photometry, Geometry, Degradation) to conserve space.}
\label{tab:attack_settings}
\begin{tabular}{l p{0.75\textwidth}}
\toprule
\textbf{Category} & \textbf{Method \& Parameters} \\
\midrule

\multirow{3}{*}{\textbf{Image Processing}} 
& \textbf{Photometry}: Brightness ($\times 6.0$), Contrast ($\times 12.0$)\\
& \textbf{Geometry}: Resize ($s=0.5$),  Center Crop($s=0.4$), Random Crop($s=0.4$)\\
& \textbf{Degradation}: Gaussian Noise ($\sigma=0.25$), Blur ($radius=1.5$), JPEG ($Q=50$)\\
\midrule

\multirow{2}{*}{\textbf{Reconstructive}} 
& \textbf{VAE Compression}: VAE-B \cite{balle2018variational} ($Q=1$), VAE-C \cite{cheng2020learned} ($Q=1$) \\
& \textbf{Generative Attack}: Regen-Diff \cite{zhao2024invisible} (SD v1.5, Strength $S=0.2$), Rinsing ($N=2, 4$) \\
\midrule

\multirow{1}{*}{\textbf{Adversarial}} 
& \textbf{Adv-Emb}: PGD-based Latent Attack \cite{an2024waves} \\
\midrule

\multirow{3}{*}{\textbf{Forgery}} 
& \textbf{Imprinting}: Latent optimization via inversion \cite{muller2025black} \\
& \textbf{Reprompting}: Inverse the image to initial latent and regeneration \cite{muller2025black}\\
& \textbf{Average}: Estimate watermark pattern by averaging residuals \cite{yang2024can}\\

\bottomrule
\end{tabular}
\vspace{-0.5em}
\end{table}

\begin{figure*}[t] 
    \centering
    \begin{subfigure}[b]{0.19\linewidth}
        \centering
        \includegraphics[width=\linewidth]{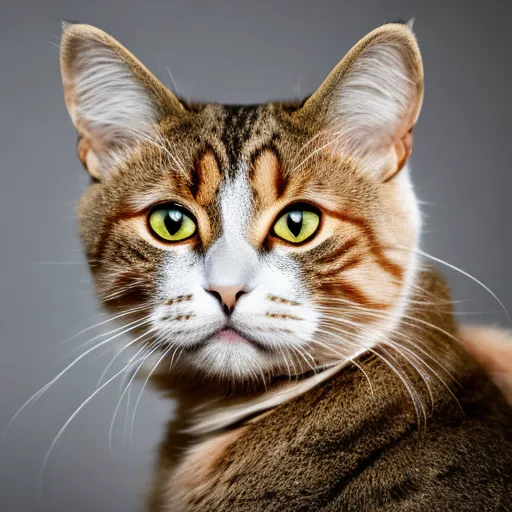}
        \caption{Original}
        \label{fig:original}
    \end{subfigure}
    \hfill
    \begin{subfigure}[b]{0.19\linewidth}
        \centering
        \includegraphics[width=\linewidth]{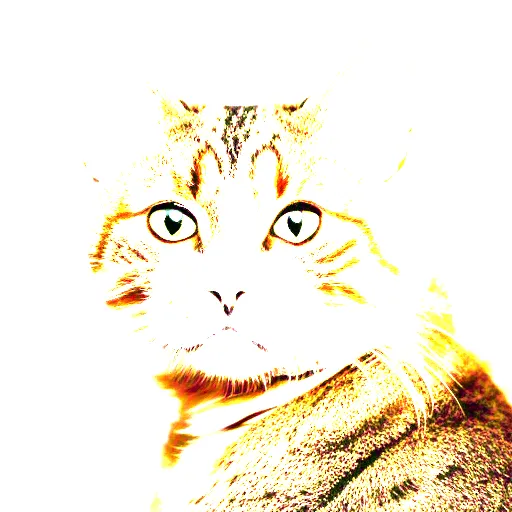}
        \caption{Brightness}
        \label{fig:brightness}
    \end{subfigure}
    \hfill
    \begin{subfigure}[b]{0.19\linewidth}
        \centering
        \includegraphics[width=\linewidth]{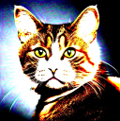}
        \caption{Contrast}
        \label{fig:contrast}
    \end{subfigure}
    \hfill
    \begin{subfigure}[b]{0.19\linewidth}
        \centering
        \includegraphics[width=\linewidth]{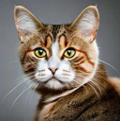}
        \caption{JPEG}
        \label{fig:jpeg}
    \end{subfigure}
    \hfill
    \begin{subfigure}[b]{0.19\linewidth}
        \centering
        \includegraphics[width=\linewidth]{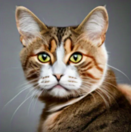}
        \caption{Blur}
        \label{fig:blur}
    \end{subfigure}
    
    \vspace{1em} 
    
    \begin{subfigure}[b]{0.19\linewidth}
        \centering
        \includegraphics[width=\linewidth]{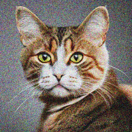}
        \caption{Noise}
        \label{fig:noise}
    \end{subfigure}
    \hfill
    \begin{subfigure}[b]{0.19\linewidth}
        \centering
        \includegraphics[width=\linewidth]{{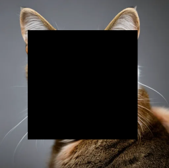}} 
        \caption{Center Crop}
        \label{fig:cc}
    \end{subfigure}
    \hfill
    \begin{subfigure}[b]{0.19\linewidth}
        \centering
        \includegraphics[width=\linewidth]{{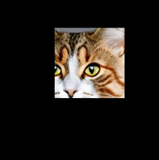}}
        \caption{Random Crop}
        \label{fig:rc}
    \end{subfigure}
    \hfill
    \begin{subfigure}[b]{0.19\linewidth}
        \centering
        \includegraphics[width=\linewidth]{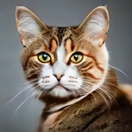}
        \caption{VAE-B}
        \label{fig:vaeb}
    \end{subfigure}
    \hfill
    \begin{subfigure}[b]{0.19\linewidth}
        \centering
        \includegraphics[width=\linewidth]{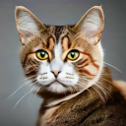}
        \caption{VAE-C}
        \label{fig:vaec}
    \end{subfigure}
    
    \caption{\textbf{Visual Robustness Examples.} Comparison of the watermarked image under various attacks: (a) Original, (b) Brightness, (c) Contrast, (d) JPEG Compression, (e) Gaussian Blur, (f) Gaussian Noise, (g) Center Crop (C.C.), (h) Random Crop (R.C.), (i) VAE Compression (BMSHJ), and (j) VAE Compression (Cheng).}
    \label{fig:visual_attacks}
    \vspace{-0.5em}
\end{figure*}

\begin{table}[t]
    \centering
    \small
    \caption{\textbf{Breakdown of Performance Gains.} We compare ALIEN against the best-performing baseline (Best SOTA) for each metric/condition. 
    \textbf{Panel A} calculates the average relative improvement across 5 quality metrics (33.1\%). 
    \textbf{Panel B} highlights the robustness gain across 15 distinct conditions (12 Generative + 3 Stability). The final 14.0\% is the weighted average gain over these 15 conditions.}
    \label{tab:gain_analysis}
    \resizebox{\textwidth}{!}{%
    \begin{tabular}{l|l|c|c|c|r}
        \toprule
        \multicolumn{6}{c}{\textbf{Panel A: Quality Improvement (ALIEN-Q vs. Best SOTA)}} \\
        \midrule
        \textbf{Metric} & \textbf{Description} & \textbf{Best SOTA} & \textbf{Method} & \textbf{ALIEN-Q} & \textbf{Improvement} \\
        \midrule
        FID $\downarrow$ & Fréchet Inception Distance & 24.42 (G.S.) & Lower is better & 24.29 & +0.5\% \\
        PSNR $\uparrow$ & Peak Signal-to-Noise Ratio & 29.09 (StableSig.) & Higher is better & 32.41 & +11.4\% \\
        SSIM $\uparrow$ & Structural Similarity & 0.922 (ZoDiac) & Higher is better & 0.949 & +2.9\% \\
        SIFID $\downarrow$ & Single Image FID & 0.105 (StableSig.) & Lower is better & 0.023 & \textbf{+78.1\%} \\
        DreamSim $\downarrow$ & Perceptual Similarity & 0.011 (StableSig.) & Lower is better & 0.003 & \textbf{+72.7\%} \\
        \midrule
        \multicolumn{5}{r|}{\textit{Average Quality Improvement}} & \textbf{+33.1\%} \\
        \bottomrule
        \toprule
        \multicolumn{6}{c}{\textbf{Panel B: Robustness Improvement (ALIEN-R vs. Training-free SOTA)}} \\
        \midrule
        \textbf{Condition Category} & \textbf{Specifics} & \textbf{Best SOTA} & \textbf{Metric} & \textbf{ALIEN-R} & \textbf{Gain} \\
        \midrule
        1. Generative Variant & Average Acc. across 12 conditions & \multirow{2}{*}{$\sim$0.84} & \multirow{2}{*}{Acc.} & \multirow{2}{*}{$\sim$0.90} & \multirow{2}{*}{+6.5\%} \\
        (12 conditions) & (4 Schedulers $\times$ Regen/Rinse) & & & & \\
        \midrule
        2. Sampler Stability & Average Acc. across 3 stochastic samplers & \multirow{2}{*}{$\sim$0.56} & \multirow{2}{*}{Acc.} & \multirow{2}{*}{$\sim$1.00} & \multirow{2}{*}{\textbf{+44.0\%}} \\
        (3 conditions) & (DPM++ SDE, Euler a, DPM2 a) & & & & \\
        \midrule
        \multicolumn{5}{r|}{\textit{Calculation: } $(12 \times 6.5\% + 3 \times 44.0\%) / 15$} & \multirow{2}{*}{\textbf{+14.0\%}} \\
        \multicolumn{5}{r|}{\textbf{Average Robustness Improvement}} & \\
        \bottomrule
    \end{tabular}%
    }
    \vspace{-0.4em}
\end{table}

\begin{table*}[t]
    \centering
    \caption{Comparison of Imperceptibility and Robustness.
    We compare 48-bit and 128-bit payloads. 
    \textit{Fidelity} is reported as Mean $\pm$ Std. Dev. 
    \textit{Robustness} is reported as Detection Accuracy.
    Abbreviations: \textit{No Att.} (No Attack), \textit{Comp.} (Compression), \textit{Comb.} (Combined Attack).}
    \label{tab:compact_comparison}
    
    \footnotesize 
    \setlength{\tabcolsep}{4pt} 
    
    \begin{tabular}{lccccccccccccc}
        \toprule
        \multirow{2}{*}{\textbf{Payload}} & \multicolumn{3}{c}{\textbf{Imperceptibility Metrics}} & \multicolumn{10}{c}{\textbf{Robustness (Bit Accuracy)}} \\
        \cmidrule(lr){2-4} \cmidrule(lr){5-14}
         & PSNR($\uparrow$) & SSIM($\uparrow$) & LPIPS($\downarrow$) & No Att. & Blur & Noise & JPEG & Resize & Sharp & Bright & Contr. & Sat. & Comb. \\
        \midrule
        
        \textbf{48 Bits} & 
        $31.17_{\pm 0.90}$ & $0.62_{\pm 0.03}$ & $0.097_{\pm 0.01}$ & 
        0.99 & 
        0.99 & 
        0.92 & 
        0.99 & 
        0.99 & 
        0.99 & 
        0.99 & 
        0.99 & 
        0.99 & 
        0.87 \\
        
        \textbf{128 Bits} & 
        $30.79_{\pm 0.85}$ & $0.59_{\pm 0.03}$ & $0.151_{\pm 0.01}$ & 
        0.99 & 
        0.99 & 
        0.90 & 
        0.98 & 
        0.99 & 
        0.99 & 
        0.99 & 
        0.99 & 
        0.99 & 
        0.82 \\
        
        \bottomrule
    \end{tabular}
    \vspace{-0.4em}
\end{table*}

\newcommand{\risk}{\textcolor{red}{$\times$}} 
\newcommand{\safe}{\textcolor{green}{\checkmark}}   

\begin{table*}[t]
    \centering
    \caption{Impact of Fixed Thresholding on False Positive Rates (ROBIN Scheme).
    Comparison of clean image means ($\mu_{clean}$) versus the optimal threshold for Cropping ($\tau_{crop}$) using the ROBIN watermarking method. 
    For both SD v1.5 and v2.1, benign degradations like Blurring and JPEG shift $\mu_{clean}$ below $\tau_{crop}$ (marked with \risk).}
    \label{tab:threshold_shift}
    
    \setlength{\tabcolsep}{5pt}
    \renewcommand{\arraystretch}{1.1}
    
    \begin{tabular}{lcccccc}
        \toprule
        \multirow{2}{*}{\textbf{Attack}} & \multicolumn{3}{c}{\textbf{Stable Diffusion v1.5} ($\tau_{crop}=39.45$)} & \multicolumn{3}{c}{\textbf{Stable Diffusion v2.1} ($\tau_{crop}=55.80$)} \\
        \cmidrule(lr){2-4} \cmidrule(lr){5-7}
         & \textbf{Clean Mean} & \textbf{Opt. $\tau$} & \textbf{Fixed $\tau$ Risk} & \textbf{Clean Mean} & \textbf{Opt. $\tau$} & \textbf{Fixed $\tau$ Risk} \\
        \midrule
        No Attack    & 40.08 & 38.28 & \safe & 55.90 & 52.74 & \safe \\
        Cropping     & 40.25 & 39.45 & -     & 55.86 & 55.80 & -     \\
        Blurring     & 32.38 & 30.31 & \risk & 52.65 & 50.24 & \risk \\
        JPEG         & 37.22 & 36.91 & \risk & 55.65 & 53.57 & \risk \\
        Color Jitter & 36.91 & 36.85 & \risk & 54.04 & 51.39 & \risk \\
        Noise        & 41.63 & 40.02 & \safe & 57.07 & 55.62 & \safe \\
        \bottomrule
    \end{tabular}
    \vspace{-1.4em}
\end{table*}

\subsection{Hyperparameters and Training Settings}
\label{app:training_details}

\noindent\textbf{Watermark Generation Module Training.} 
We train the Imperceptible Latent Watermark Generation module using the AdamW optimizer with a learning rate of $5 \times 10^{-5}$ and a weight decay of $1 \times 10^{-5}$. 
We construct a synthetic training set of 10,000 images generated based on COCO2014 \cite{lin2014microsoft} prompts. The model is trained for 50,000 steps with a batch size of 16.
To ensure training stability and perceptual quality, we impose a maximum gradient norm of $10^{-2}$. 
The total objective combines the secret recovery loss ($\mathcal{L}_{sec}$), pixel-wise reconstruction loss ($\mathcal{L}_{mse}$), and perceptual loss ($\mathcal{L}_{lpips}$ using AlexNet backbone). 
The loss weights are set to $\lambda_{sec}=1.0$, $\lambda_{mse}=30.0$, and $\lambda_{lpips}=0.3$. 
Notably, $\lambda_{mse}$ and $\lambda_{lpips}$ are linearly ramped up from 0 to their peak values over the first 5,000 steps to facilitate stable convergence in the early training phase. 

\noindent\textbf{Injection Settings.} 
During the inference phase, ALIEN-Q injects watermarks within the sampling interval of steps 20--45 (with injection strength $\lambda=0.85$), while ALIEN-R extends the injection window to cover steps 0--50 ($\lambda=1.0$) for enhanced robustness.

\section{Experimental Setup and Extended Experiments}
\label{app:setup}

\subsection{Dataset Settings}
\label{app:datasets}

\noindent\textbf{Evaluation Datasets.} We utilize two distinct datasets to comprehensively evaluate both watermarking performance and image generation quality: For measuring detection accuracy (TPR@1\%FPR) and payload capacity (Bit Accuracy), we employ the Stable Diffusion-Prompts (SDP) dataset \cite{Gustavosta2023}. We randomly select 350 prompts from the dataset to generate image pairs (watermarked vs. clean) for distinct evaluation. For evaluating generative fidelity, specifically Fréchet Inception Distance (FID), we utilize the MS-COCO dataset \cite{lin2014microsoft}. We randomly sample 5,000 captions from the MS-COCO validation set to generate 5,000 watermarked images and compute the FID score against the corresponding real reference images to ensure standardized quality comparison.

\noindent\textbf{Generation Configuration.} We conduct experiments on two representative Latent Diffusion Models: Stable Diffusion v1.5 and Stable Diffusion v2.1. Unless otherwise specified, we use the DDIM sampler with 50 inference steps. The classifier-free guidance scale is set to 7.5. All generated images are of resolution $512 \times 512$. All experiments were conducted on a single NVIDIA RTX 3090 GPU.

\subsection{Robustness Evaluation Settings}
\label{app:robustness_settings}

\subsection{Extended Experiments}
\label{app:add_experiments}

\paragraph{Detailed Performance Gain Analysis.}
To substantiate the claims made in the abstract, we provide a detailed breakdown of the performance improvements. Table~\ref{tab:gain_analysis} illustrates the calculation of the \textbf{33.1\% quality improvement} and \textbf{14.0\% robustness improvement} compared to the state-of-the-art (SOTA) methods.


\paragraph{Impact of Payload Size.} We further investigate the impact of payload size on the trade-off between watermark capacity, imperceptibility, and robustness. As demonstrated in Table \ref{tab:compact_comparison}, increasing the payload capacity from 48 bits to 128 bits results in only a minor decrease in fidelity metrics, exemplified by a slight PSNR drop from 31.17 dB to 30.79 dB. This result validates that ALIEN effectively supports high-capacity embedding while maintaining visual quality comparable to low-capacity settings.


\paragraph{Threshold Determination for Distance-Based Methods.}
We evaluate the impact of distribution shifts caused by image degradations on threshold determination for the distance-based method, ROBIN. As shown in Table \ref{tab:threshold_shift}, using a fixed threshold derived from a challenging scenario (Cropping, $\tau=39.45$) poses risks when applied to other distortions. We observe a clear distribution shift in the metric space for degradations that suppress high-frequency information. For instance, the mean metric of unwatermarked images under Blurring drops to 32.38, falling below the fixed cropping threshold of 39.45. Since detection occurs when the metric is below the threshold, this shift results in increased False Positive Rates (FPR), where benign, low-quality images are misclassified. 

\section{Extended Related Work}
\label{app:related}

This section provides a detailed review and taxonomy of existing watermarking methods for generative models.

\subsection{Watermarking Schemes Taxonomy}
\label{app:taxonomy}

We categorize existing watermarking methods for generative models into five primary classes based on their embedding stage: Random Seed Modification, U-Net Modification, VAE Modification, Latent Space Modification, and Post-Processing.

\textbf{ Random Seed Modification (Initial Latent).} 
Methods in this category embed watermarks by changing in the sampling process of the initial latent variables. For instance, \textbf{Tree-Ring} \cite{wen2024tree} modifies the initial noise vector in the Fourier domain to embed a ring-shaped pattern, which is detectable via diffusion inversion. \textbf{Gaussian Shading} \cite{yang2024gaussian} employs a constrained sampling strategy to apply specific patterns to the initial latent.

\textbf{ Model Modification.} 
These approaches embed watermarks by fine-tuning specific components of the generative model to ensure watermark preservation during generation. 
\textbf{Stable Signature} \cite{fernandez2023stable} fine-tunes the VAE decoder to embed the watermark into the pixel space during the latent-to-image decoding stage. \textbf{Aqualora} \cite{feng2024aqualora} modifies the U-Net to inject the watermark during the iterative denoising process.

\textbf{ Latent Space Optimization.} 
Unlike static encoding methods, these approaches formulate watermark embedding as an optimization problem within the latent space. 
\textbf{Zodiac} \cite{zhang2024attack} embeds the watermark by iteratively optimizing the \textit{initial} latent variable to ensure high detectability in the generated output. 
\textbf{ROBIN} \cite{huang2024robin} focuses on optimizing the \textit{intermediate} latent representations during the diffusion process. It incorporates learnable prompts to align the watermarked latents with the text condition.

\textbf{Post-Processing Methods.} 
These techniques apply digital watermarking algorithms to the image after it has been fully generated, operating independently of the generation pipeline. \textbf{StegaStamp} \cite{tancik2020stegastamp} is a deep learning-based encoder-decoder framework that embeds invisible hyperlinks or bit-strings into the final image output.

\end{document}